\documentclass[a4paper,11pt]{article}
\pdfoutput=1 

\usepackage{jheppub} 

\usepackage[T1]{fontenc} 
\usepackage{caption}
\usepackage{tikz}
\usepackage{subcaption}
\usepackage{graphicx}
\usepackage{array}
\usepackage{makecell}
\usepackage{longtable}
\usepackage{bm}

\definecolor{darkgreen}{rgb}{0.1,0.5,0.1}
\definecolor{cerulean}{rgb}{0.0, 0.48, 0.65}

\newcommand{\pt}{\ensuremath{p_\textrm{T}}}
\newcommand{\pT}{\pt}

\newcommand{\rev}[1]{{#1}}

\title{Resolving Extreme Jet Substructure}

\author{Yadong Lu$^{a}$}
\author{Alexis Romero$^{b}$}
\author{Michael James Fenton$^{b}$}
\author{Daniel Whiteson$^{b}$}
\author{Pierre Baldi$^{c}$}
\emailAdd{yadongl1@uci.edu}
\emailAdd{alexir2@uci.edu}
\emailAdd{m.fenton@uci.edu}
\emailAdd{daniel@uci.edu}
\emailAdd{pfbaldi@uci.edu}

\affiliation{$^{a}$Department of Statistics, University of California, Irvine, CA, USA 92627}
\affiliation{$^{b}$Department of Physics and Astronomy, University of California, Irvine, CA, USA 92627}
\affiliation{$^{c}$Department of Computer Science, University of California, Irvine, CA, USA 92627}

\abstract{ We study the effectiveness of theoretically-motivated high-level jet observables in the extreme context of jets with a large number of hard sub-jets (up to $N=8$).  Previous studies indicate that high-level observables are powerful, interpretable tools to probe jet substructure for $N\le 3$ hard sub-jets, but that deep neural networks trained on low-level jet constituents match or slightly exceed their performance. We extend this work for up to $N=8$ hard sub-jets, using deep particle-flow networks (PFNs) and Transformer based networks to estimate a loose upper bound on the classification performance.  A fully-connected neural network operating on a standard set of high-level jet observables, 135 $\textrm{N}$-subjetiness observables and jet mass, reach classification accuracy of 86.90\%, but fall short of the PFN and Transformer models, which reach classification accuracies of 89.19\% and 91.27\% respectively, suggesting that the constituent networks utilize information not captured by the set of high-level observables. We then identify additional high-level observables which are able to narrow this gap, and utilize LASSO regularization for feature selection to identify and rank the most relevant observables and provide further insights into the learning strategies used by the constituent-based neural networks.  The final model contains only 31 high-level observables and is able to match the performance of the PFN and approximate the performance of the Transformer model to within 2\%.
}

\begin{document} 
\maketitle
\flushbottom

\section{\label{sec:level1}Introduction}

The era of the Large Hadron Collider (LHC) has opened a new frontier in jet physics: the interior of jets.  Hadronically decaying objects with large transverse momentum produce collimated quarks and gluons which lead to jets with complex energy patterns, including some with multiple, distinct sub-jets~\cite{Butterworth:2008iy}. Jets with two or three sub-jets have been observed and extensively studied~\cite{Hook:2012fd,Aaboud:2018psm,Khachatryan:2014vla, Aaboud:2019aii, Aad:2020zcn, Sirunyan:2018asm, Sirunyan_2020}, and jets with additional hard subjets will become more important as the high-luminosity LHC  collects large datasets in which high-$p_\textrm{T}$ objects appear in greater numbers~\cite{CMS:2021beq,Aad:2020ddw}. Additionally, future colliders may push the energy frontier forward, creating jets with many more hard sub-jets. In these settings, distinguishing jets with multiple hard sub-jets will be an increasingly important element in searches for new physics. 

Many theoretically-motivated jet substructure techniques have been proposed and studied to identify jets with multiple hard sub-jets by summarizing the information of the low-level jet constituents into a compact set of high-level observables~\cite{ Komiske:2017aww,Larkoski:2013eya}.
However, recent strides in deep learning have demonstrated the ability to extract information directly from low-level detector data, making them powerful probes of the information content of the jets~\cite{Komiske:2016rsd,Baldi:2014kfa,Almeida:2015jua,deOliveira:2015xxd,Baldi:2016fql,Larkoski:2017jix,Kasieczka:2017nvn,baldi_2021}.  Application of these deep neural networks (DNNs) to jets has revealed that there is often additional information available in the constituents that is not captured by such high-level observables.  

Specifically, studies with two hard sub-jets~\cite{Baldi:2016fql,Faucett:2020vbu} have found a small but stubborn gap when comparing the performance of DNNs on low-level jet constituents to networks on high-level observables, indicating that the observables fail to capture all of the information used by the constituent-based networks. In studies of jets with three hard sub-jets, several studies have analyzed the performance gap between low- and high-level jet information in the context of top-quark tagging. While some have found a negligible gap between high-level observables and pixelated jet image representations \cite{Moore:2018lsr}, modern deep learning architectures operating on unpixelated low-level constituent inputs continue to outperform high-level networks~\cite{Kasieczka:2019dbj, Qu:2019gqs, Mikuni:2021pou,CMS:2021mjl}. For jets with four hard subjets,  taggers have been developed which use high-level observables~\cite{Aguilar-Saavedra:2020uhm,Aguilar-Saavedra:2017rzt,Chen:2020ywp}, but without a comparison to the performance of networks using low-level jet information.

In this paper, we extend jet tagging studies to jets with up to eight hard sub-jets, to probe the question of whether the existing high-level observables are sufficient to analyze such extreme jets, or whether modern deep learning architectures can identify additional, untapped information. In each case, we compare the performance of networks which use compact high-level observables to those that take voluminous low-level calorimeter data.\footnote{A similar strategy may be applied to tracking information, which is left for future studies.} We then use the networks trained on low-level data as a probe, and attempt to map their strategies into a set of high-level observables that contain the full range of discriminating information.

The rest of this paper is structured as follows. Section~\ref{sec:data} contains the details of the dataset generation, and describes the different topologies used to simulate jets with many hard sub-jets.  Section~\ref{sec:hl} describes the fully-connected network operating on existing high-level observables, followed by Sec.~\ref{sec:ll} which describes the networks applied to the low-level jet constituents.  Section~\ref{sec:perf} compares the performance of the network based on high-level observables to the constituent-based networks. In Sec.~\ref{sec:int}, we study the performance gap between the various networks, and find additional high-level observables which narrow the gap to the constituent-based networks. Concluding remarks are given in Sec.~\ref{sec:conc}.

\section{Dataset and Preprocessing}
\label{sec:data}

\begin{figure}[htb]
    \centering
     \begin{subfigure}[b]{0.3\textwidth}
         \includegraphics[width=4cm]{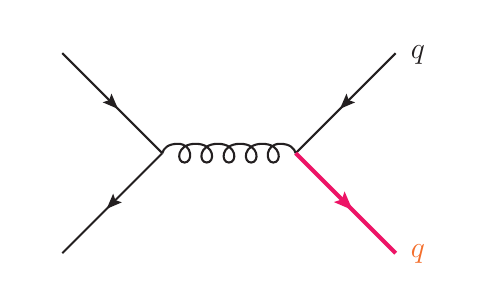}
         \caption{$N=1$}
         \label{diag_n1}
     \end{subfigure}
     \begin{subfigure}[b]{0.3\textwidth}
        \includegraphics[width=5.5cm]{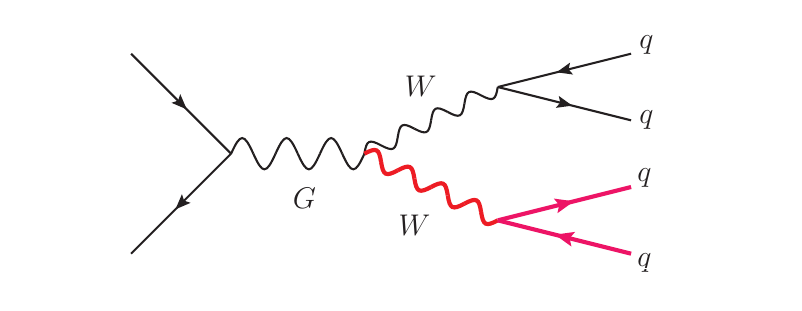}
        \caption{$N=2$}
        \label{diag_n2}
     \end{subfigure}
     \begin{subfigure}[b]{0.3\textwidth}
        \includegraphics[width=5cm]{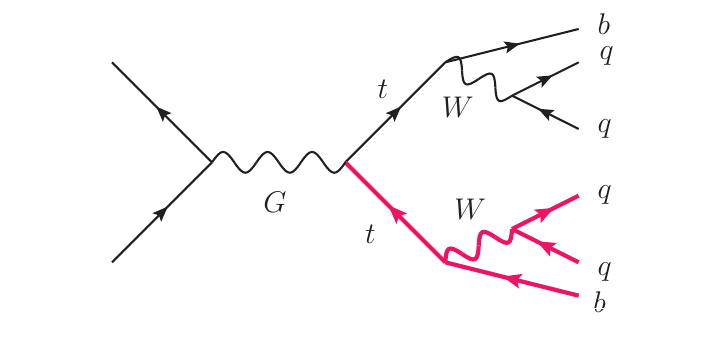}
        \caption{$N=3$}
        \label{diag_n3}
     \end{subfigure}
     \begin{subfigure}[b]{0.3\textwidth}
        \includegraphics[width=4.5cm]{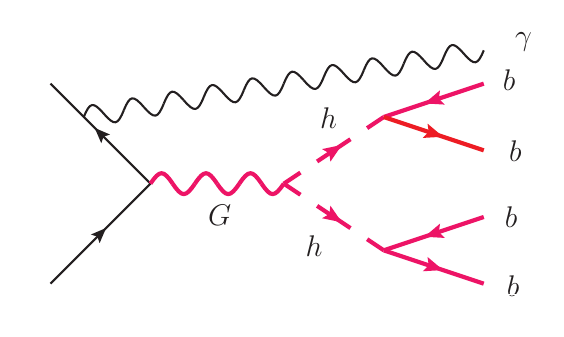}
        \caption{$N=4b$}
        \label{diag_n4}
     \end{subfigure}
     \begin{subfigure}[b]{0.3\textwidth}
        \includegraphics[width=4.5cm]{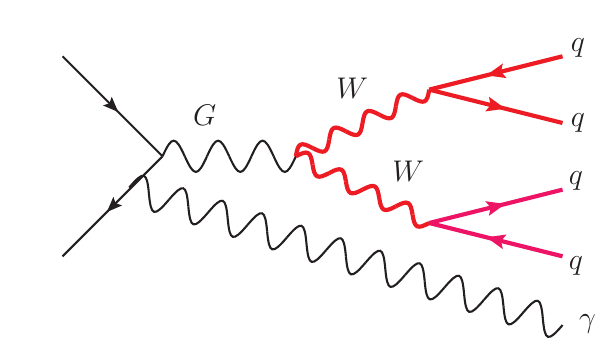}
        \caption{$N=4q$}
        \label{diag_n4q}
     \end{subfigure}
     \\
     \begin{subfigure}[b]{0.3\textwidth}
        \includegraphics[width=5cm]{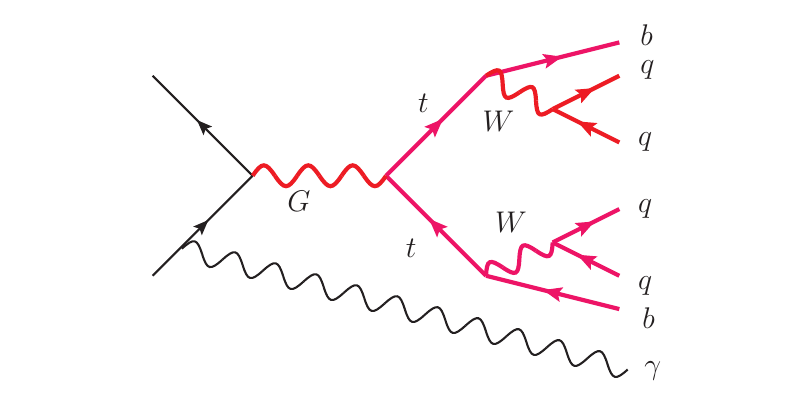}
        \caption{$N=6$}
        \label{diag_n6}
     \end{subfigure}
     \begin{subfigure}[b]{0.3\textwidth}
        \includegraphics[width=5cm]{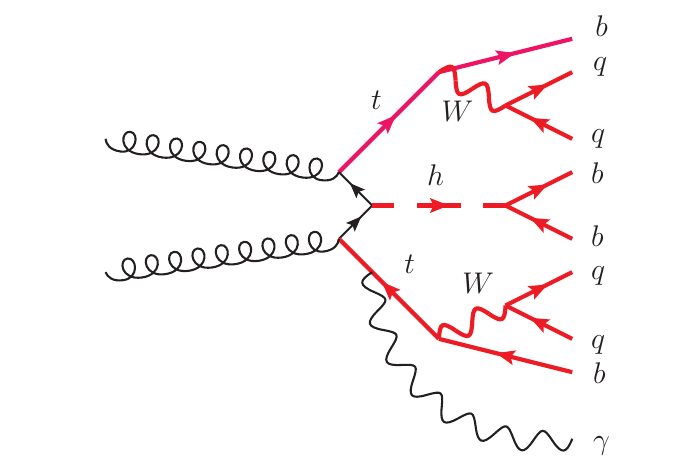}
        \caption{$N=8$}
        \label{diag_n8}
     \end{subfigure}
    \caption{Feynman diagrams for  processes which generate jets with $N=1,2,3,4b, 4q, 6,8$ hard sub-jets. The lines in red indicate the components which are required to be truth-matched to the jet. See text and  Table~\ref{tab:proc} for further generation details.}
    \label{fig:diagrams}
\end{figure}

Simulated proton-proton collision events enriched in jets with many collimated quarks are generated using the processes shown in Figure~\ref{fig:diagrams}. Most samples use the decay of a hypothetical heavy particle, such as a graviton ($G$), which subsequently decays via heavy Standard Model (SM) particles such as $W$ bosons, Higgs bosons, or top quarks, whose hadronic decays yield collimated pairs and triplets of quarks that contribute $N=2$ hard sub-jets per boson or $N=3$ hard sub-jets per top quark. For processes with $N \geq 4$ hard sub-jets, events are further required to contain high-energy photon radiation which can, for example, boost a $G\rightarrow t\bar{t}$ decay into a single jet with $N=6$ hard sub-jets.

Samples are generated for $N=1,2,3,4,6,8$ hard sub-jet classes. To probe the dependence of the classification on the topology of the decay, the $N=4$ class is subdivided in two: the $N=4q$ case, in which two collimated $W$ bosons produce a jet with four hard sub-jets from four light quarks, and the $N=4b$ case, in which two collimated Higgs bosons result in a jet with four hard sub-jets from four $b$-quarks. 
A summary of the generated samples is given in Table~\ref{tab:proc}.

\begin{table}[!h]
    \centering
        \caption{ Processes used to generate jets with $N=1,2,3,4b,4q,6,8$ hard sub-jets. Also shown are the particle masses and generator-level requirements used to more efficiently produce jets with $p_\textrm{T} \in [1000, 1200]$ GeV and mass $\in [300,700]$ GeV. All masses and momenta are in GeV. Selected $W$ boson masses of (264.5,440.8,617.1) correspond to $M_Z = (300,500,700)$, respectively. See Figure~\ref{fig:diagrams} for the corresponding Feynman diagrams.}
        \label{tab:proc}
        \begin{tabular}{c|l|rrrr|l}
\hline\hline
$N$ hard sub-jets & Process & $M_W$ & $M_h$ & $M_t$ & $M_G$ & requirements \\
\hline
1     &  $q\bar{q}\rightarrow q\bar{q}$ & & & & & $p_\textrm{T}^q > 1000$\\
2     &  $q\bar{q}\rightarrow G \rightarrow W^+W^-$ & $80.4$  & & &  $2200$ & \\
    &   & $264.5$  & & &  $2200$  & \\
        &   & $440.8$  & & &  $2500$ & \\
    &   & $617.1$  & & &  $2800$ & \\
3 &  $q\bar{q}\rightarrow G \rightarrow t\bar{t}$ &  & & 300  & 2200  & \\
  &  & & & 500  & 2500  & \\
    &  & & & 700  & 3000  & \\
4$b$ &     $q\bar{q}\rightarrow \gamma G  \rightarrow \gamma h h$ & & & & 400  &  $p_\textrm{T}^\gamma > 1000$\\
  & & & & & 600  & $p_\textrm{T}^\gamma > 1000$\\
  & & & & & 800  & $p_\textrm{T}^\gamma > 1000$\\
  4$q$ &     $q\bar{q}\rightarrow \gamma G \rightarrow \gamma W^+W^-$ & & & & 400  &  $p_\textrm{T}^\gamma > 1000$\\
  & & & & & 600  & $p_\textrm{T}^\gamma > 1000$\\
  & & & & & 800  & $p_\textrm{T}^\gamma > 1000$\\
6 &  $q\bar{q}\rightarrow \gamma G \rightarrow \gamma t\bar{t}$ &  & &  & 400  & $p_\textrm{T}^\gamma > 1000$\\
 & & & &  & 600  & $p_\textrm{T}^\gamma > 1000$\\
 & & & &  & 800  & $p_\textrm{T}^\gamma > 1000$\\
8 &  $q\bar{q}\rightarrow \gamma t\bar{t} h$ & & 100  & 125  & & $p_\textrm{T}^\gamma > 1000$\\
& & & 125  & 175  & & $p_\textrm{T}^\gamma > 1000$\\
\hline\hline
\end{tabular}
    
\end{table}

Proton-proton collisions at a center-of-mass energy $\sqrt{s}=13~$TeV are simulated with {\sc Madgraph5} v2.8.1~\cite{madgraph}, showered and hadronized with {\sc Pythia} 8.244~\cite{pythia}, and the detector response is simulated with {\sc Delphes} 3.4.2~\cite{delphes} using an ATLAS-like card with calorimeter grids of uniform width $0.0125$ in both $\eta$ and $\phi$. This fine granularity ensures that our studies include some of the effects of detector respond, but
probe the limits of the algorithms rather than the resolution of the detector. Jets are clustered using the anti-$k_{\textrm{T}}$ algorithm~\cite{Cacciari:2008gp} with radius parameter $R = 1.2$ using \textsc{FastJet 3.1.2}~\cite{Cacciari:2011ma}. Only jets with $p_\textrm{T}$ in the range of $[1000, 1200]$ GeV and jet mass in the range of $[300, 700]$ GeV are kept. The quarks produced from each process are truth-matched to the large-radius jet by requiring $\Delta R = \sqrt{\Delta\phi^2 + \Delta\eta^2} < 1.2$; only jets with the full set of quarks passing this requirement are kept. 

\begin{figure*}[!h]
    \centering
    \includegraphics[width=1\linewidth]{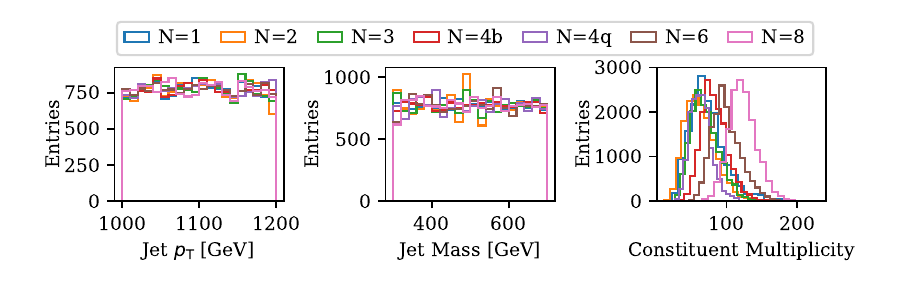}
    \caption{ 
    Distributions of jet \pT~, mass, and constituent multiplicity for the simulated jets with $N=1,2,3,4b,4q,6,8$ hard sub-jets.}
\label{fig:masspt}
\end{figure*}

To generate jets with a variety of masses, several choices are made for the intermediate particle masses. The resulting spectrum of generated jet masses features clear artifacts due to these choices.  Similarly, the distribution of jet $p_\textrm{T}$ shows some dependence on the process used to generate the set of collimated quarks.  To avoid learning artifacts in jet mass and $p_\textrm{T}$, we selectively reject events until we achieve uniform distributions of jet  mass, $p_\textrm{T}$, and $N$ hard sub-jets. This process yields a balanced sample of unweighted events, at the expense of reduced generation efficiency.  The balanced histograms are shown in Figure~\ref{fig:masspt}, demonstrating an approximately uniform distribution in both jet mass and \pT. 

Since the low-level networks require fixed-size arrays as input, the events are zero-padded with zero-$\pT$ particles to ensure all input arrays have a length of 230. We find that padding the events to this length is more than enough to capture all hard constituents in the events, as they have a mean constituent multiplicity of 82; see Figure~\ref{fig:masspt}.

The jets are preprocessed by normalizing the $p_\textrm{T}$ of their constituents to sum to unity, and centered based on the $p_\textrm{T}$-weighted arithmetic mean in $\eta$ and circular mean in $\phi$. In total, we have 108,359 simulated jets (around 15,480 for each $N$), which we divide into a training set, a validation set, and a test set with proportions $80:10:10$, respectively. We use 10-fold cross validation to ensure statistical robustness for all results.

\section{High-Level Observables Network}
\label{sec:hl}
\begin{figure*}[!h]
    \centering
    \includegraphics[width=\linewidth]{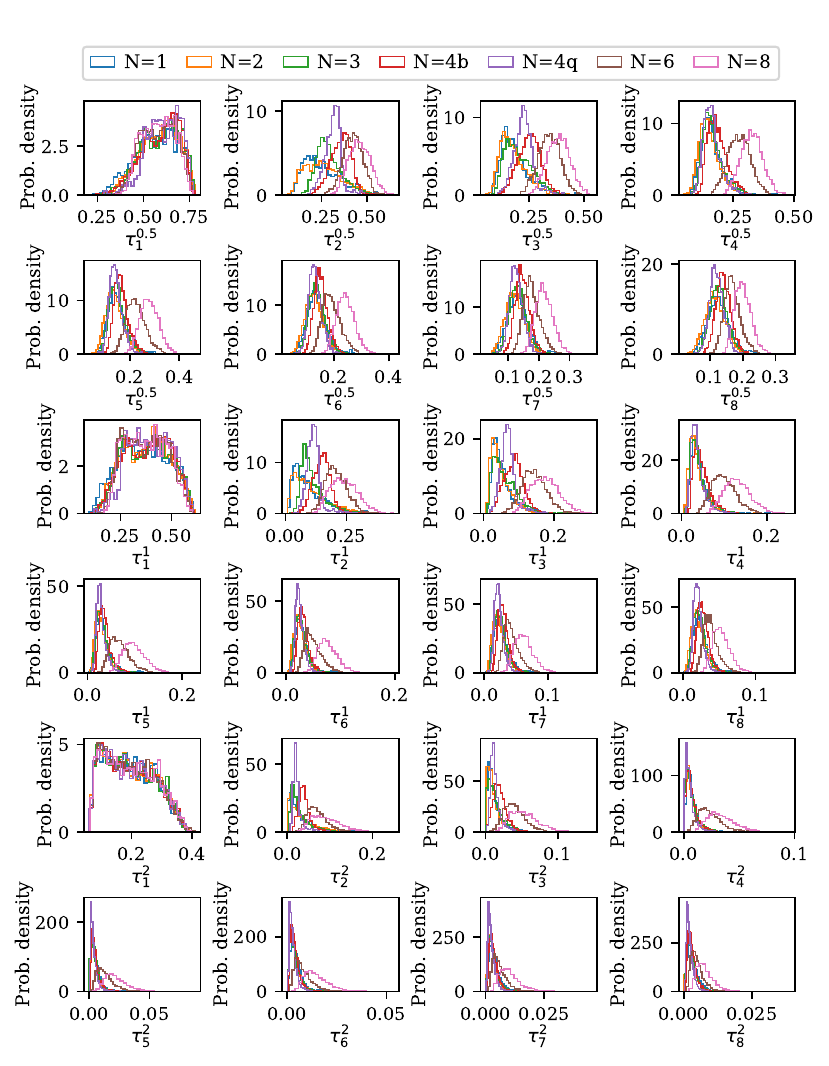}
    \caption{ Distributions of $\textrm{N}$-subjettiness~\cite{Thaler:2010tr} variables $\tau_\textrm{N}^\beta$ for $\textrm{N} = 1, \ldots, 8$, and $\beta \in \{ \frac{1}{2}, 1, 2 \}$, in samples of simulated jets with various $N$ hard sub-jets.}
    \label{fig:hl_hist_nsubs}
    \vspace{0.5cm}
\end{figure*}

Identification of jets with $N$ hard sub-jets is a well-explored topic experimentally for $N \le 3$, for which many theoretically motivated observables have been constructed to summarize the information contained in the jet energy pattern~\cite{Thaler:2010tr, Larkoski:2013eya,Komiske:2016rsd, Komiske:2017aww}. These observables have the advantage that they are compact and physically interpretable, and can in principle be applied to jets with many more hard sub-jets. Here, we use $\textrm{N}$-subjettiness~\cite{Thaler:2010tr}, together with the jet mass, as a well-known benchmark.  Following Ref.~\cite{Datta_2017}, a total of 135 $\textrm{N}$-subjettiness observables ($\tau_\textrm{N}^\beta$) are calculated along the $k_\textrm{T}$ axis, with the sub-jet axis parameter $\textrm{N}=1,\ldots, 45$, and angular weighing exponent $\beta \in \{\frac{1}{2}, 1, 2\}$, which together with the jet mass account for 136 jet observables. Distributions of some of the $\textrm{N}$-subjettiness observables are shown in Figure~\ref{fig:hl_hist_nsubs} for jets with various numbers of hard sub-jets.

We train a fully-connected (dense) network which operates on the 136 high-level observables, referred to as DNN$_{136}$. A grid search of the hyperparameters of the dense network indicates that the best structure for the network has six hidden layers of size (800-800-800-800-800-64) and \textsc{ReLu}~\cite{inproceedings} activation function. To prevent overfitting and to facilitate training stability, dropout with rate 0.3 and batch normalization are applied respectively after every hidden layer. The output layer is a 7-dimensional softmax function, with one dimension for each category of $N$ hard sub-jets. Please refer to Table~\ref{tab:ModelSummaries} for the model summary of DNN$_{136}$.

\section{Low-Level Jet Constituent Models}
\label{sec:ll}

The primary focus of our study is to answer the question of whether the patterns of energy depositions in the jets contain additional information useful to the classification task that is not captured by the high-level observables.  To probe the relevant information content, we train networks which operate directly on the low-level calorimeter tower constituents of the jets.  We focus on two such network architectures; Particle-Flow Networks (PFN)~\cite{efn} and Transformers~\cite{transformer}. Both networks have matched or outperformed other low-level network architectures, such as convolutional networks, in a variety of classification tasks~\cite{efn,Faucett:2020vbu,Collado:2021opi}, but the decision to chose PFNs and Transformers to probe the information of content of the jets relies on their specific learning strategies. \rev{PFNs learn an event-level latent representation of the jets, see Sec.~\ref{sec:pfn}. Transformers, on the other hand, learn a contextualized embedding of each constituent and use self-attention mechanisms to determine which parts of the embedded input sequence to focus on in order to make a final prediction, see Sec.~\ref{sec:transformer}. Both networks are invariant to the input ordering, which makes them well-suited for jet classification tasks.} 

\subsection{Particle-Flow Networks}
\label{sec:pfn}
The power of Particle-Flow Networks (PFNs)~\cite{efn} relies on their ability to learn virtually any symmetric function of the jet constituents.  Their mathematical structure is naturally invariant under permutation of the input ordering, constructed as a summation over the constituents:
\begin{equation}
    \textrm{PFN} : F \left( \sum_{i \in \text{jet}}  \Phi(p_{\textrm{T}i}, \eta_i, \phi_i) \right),
    \label{eq:PFN}
\end{equation}
where in our case $\Phi(p_{\textrm{T}i}, \eta_i, \phi_i)$ represents the per-tower latent space operating on the normalized three-momentum of constituent tower $i$, and $F$ represents the jet-level latent space. 

A grid search of the hyperparameters of the PFN finds that the best structure for the network has two layers in the $\Phi$ module of size $(128, 128)$ and two layers in the $F$ module of size $(1024, 1024)$, with a drop out rate of 0.2. All hidden layers have \textsc{ReLu}~\cite{inproceedings} activation functions. The output layer is the same as for DNN$_{136}$; a 7-dimensional softmax with one element for each possible value of $N$ hard sub-jets. Please refer to Table~\ref{tab:ModelSummaries} for a summary of the PFN model.

\subsection{Transformers}
\label{sec:transformer}
Like the PFN, the Transformer~\cite{transformer} model also operates on the low-level jet constituents. Transformers are characterized by the use of the self-attention mechanism, which learns to focus on the parts of the input sequence that the model deems important, to extract meaningful representations of the input sequence. With self-attention layers, Transformer models and its variants achieve state-of-the-art performance in a variety of sequence data modeling tasks~\cite{devlin2019bert, gpt2, lewis2019bart}, and in applications to particle physics tasks with inherent symmetries~\cite{Fenton:2020woz,Shmakov:2021qdz}.  In our application, we employ the Encoder model from the Transformer in~\cite{devlin2019bert}, which has a stack of multiple attention vectors computed in parallel to increase the expressiveness of the network.

A grid search of the hyperparameters of the Transformer finds that the best structure for the network has four transformer layers, with hidden size of 256 and intermediate dimension of 128. For computational efficiency and to compare networks of similar complexity (see Table~\ref{tab:ModelSummaries}), we focus our search on smaller architectures compared to the ones used in the original paper \cite{transformer, devlin2019bert}.  A summary of the Transformer model is given in Table~\ref{tab:ModelSummaries}.

\section{Performance}
\label{sec:perf}

\begin{figure*}[!h]
\label{fig:perf_n}
    \centering
    \includegraphics[width=\linewidth]{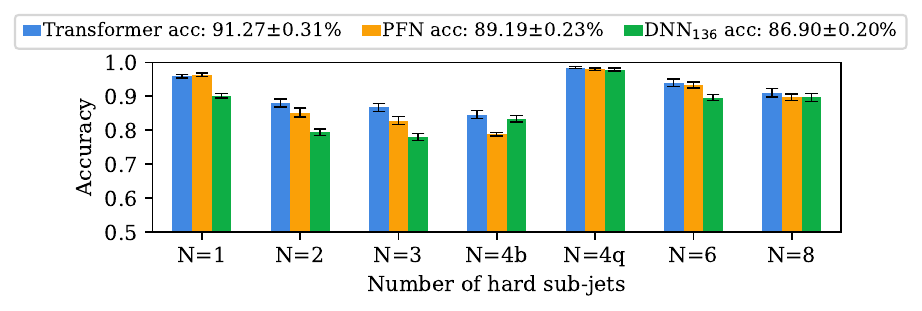}
    \caption{ 
    Mean 10-fold accuracy and statistical uncertainty of the network predictions for jets in each class, for the three network studies.}
    \label{fig:acc}
    
    \vspace{0.5cm}
    
    \centering
    \includegraphics[width=\linewidth]{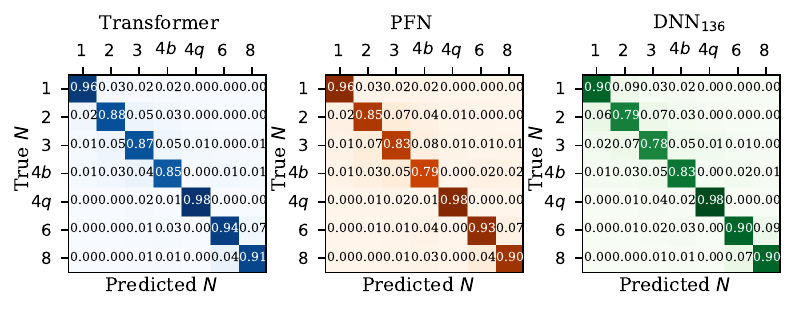}
    \caption{
    Confusion matrices of the three networks studies. The entries show the frequency at which the networks predict a jet class for a given class.}
\label{fig:cf_matrix}
\end{figure*}

We measure a network's accuracy as the fraction of correctly identified jets; specifically, the fraction of jets where the predicted class matches the true class. For jets in each class, the 10-fold average accuracy is shown in Figure~\ref{fig:acc}. The Transformer network achieves the highest overall classification performance, with an accuracy of $91.27 \pm 0.31\%$, followed by the PFN with an accuracy of $89.19 \pm 0.23\%$.\footnote{We compared the performance of the PFN to an Energy-Flow Network~\cite{efn}, which has a strictly linear dependence on \pT, enforcing IRC safety. The performance of the PFN and EFN were equivalent, indicating that the PFN was not learning IRC unsafe information.} The high-level model DNN$_{136}$ is the least performant model, with an accuracy of $86.90 \pm 0.20\%$. The accuracy of the networks is, however, not uniform across all classes, as some classes achieve better performance than others. Despite this, the relative ranking of the three networks is mostly the same for each class, with the exception of N=4b, in which DNN$_{136}$ outperforms the PFN. The classes with the lowest accuracy scores are $N=2$, $N=3$, and $N=4b$. The confusion matrices showing the mean 10-fold classification predictions for the jets in each class are shown in Figure~\ref{fig:cf_matrix}, which show that these classes are often misclassified among each other, and with $N=1$ to a lesser extent.  The confusion matrices also show a high degree of diagonality in all networks, with the largest off-diagonal elements located typically in the adjacent categories. This suggests that the networks have learned to identify the number of hard sub-jets, with the largest classification mistype in the $N\pm 1$ classes, as might be expected. It is however interesting to note how the class with the highest accuracy score for all networks is $N=4q$, and how infrequently it is misclassified with $N=4b$, suggesting that the networks are learning more information than simply the number of hard sub-jets.

\begin{figure*}[!h]
    \centering
    \includegraphics[width=\linewidth]{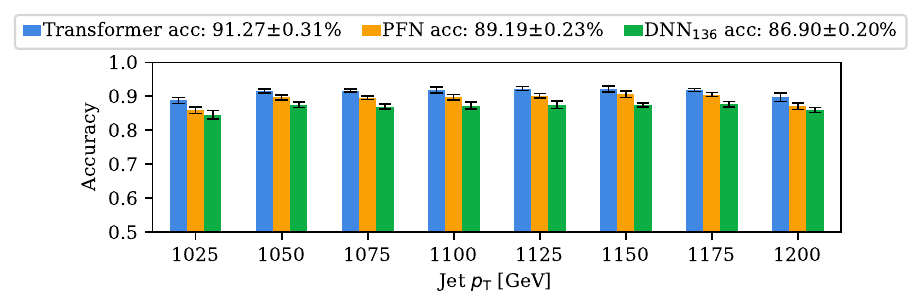}
    \caption{ 
        Mean 10-fold accuracy and statistical uncertainty  of the network prediction for jets \rev{within ranges of jet \pT}, for the three network studies. \rev{The jets are binned according to their \pT\ in intervals of 25 GeV. The x-axis labels correspond to the upper bound of the intervals.}}
    \label{fig:perf_pt}
    
    \vspace{0.1cm}

    \centering
    \includegraphics[width=\linewidth]{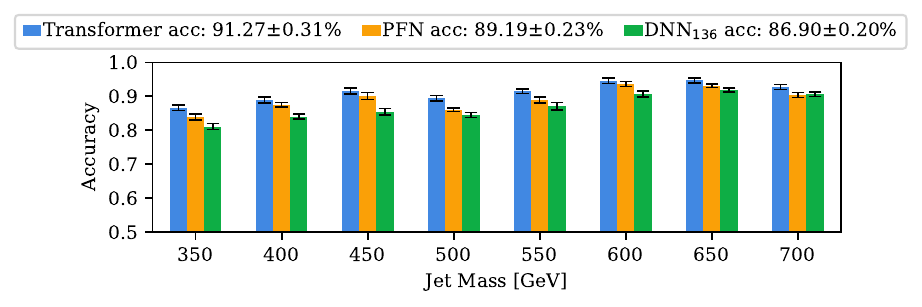}
    \caption{ 
        Mean 10-fold accuracy and statistical uncertainty of the network prediction for jets \rev{within ranges of jet mass}, for the three network studies. \rev{The jets are binned according to their mass in intervals of 50 GeV. The x-axis labels correspond to the upper bound of the intervals.}}
    \label{fig:perf_mass}
\end{figure*}

To study whether the networks have a dependence on the $p_{\textrm{T}}$ of the jets, the accuracy of their predictions as a function of jet $p_{\textrm{T}}$ is shown in Figure~\ref{fig:perf_pt}. For all networks, the accuracy remains relatively constant across the spectrum, indicating that the networks do not have a strong $p_{\textrm{T}}$ dependence, as expected given the balancing of the dataset. In all ranges, the Transformer slightly outperforms the PFN, both followed by the DNN$_{136}$. 

We also study whether the networks have a dependence on the jet mass by assessing the accuracy of their predictions as a function of jet mass,\footnote{\rev{As a crosscheck, we tested the mass dependence of the networks across smaller mass ranges, and found that their performance agrees with the results shown in Figure~\ref{fig:perf_mass}.}} as shown in Figure~\ref{fig:perf_mass}. While the Transformer outperforms the PFN and the DNN$_{136}$ in all ranges, the accuracy of the network's predictions is not uniform, but generally rises with jet mass, in contrast with the lack of variation with jet \pT. This dependence is further studied in Section~\ref{sec:ranking_obs} by inspecting the correlations between the most important high-level observables and the jet mass.

To our knowledge, this is the first comparison of high-level observables to networks which use jet constituents for jets with more than three hard sub-jets.  Our results indicate that observables such as $\textrm{N}$-subjettiness variables are powerful discriminants for jets with many hard sub-jets, and perform well when tested on jets with different topologies. There is, however, a small but persistent gap when between their overall performance and that of the low-level networks. In the following sections, we identify new observables which can be combined with the $\textrm{N}$-subjettiness variables to bridge this gap.

\section{Closing the Performance Gap}
\label{sec:int}

The Transformer and PFN models operating on the low-level jet constituents have generally outperformed the high-level observables across the different classes $N$ and across the jet mass and \pT~spectra. Low-level networks have the advantage of efficiently extracting information from the jet constituents to achieve state-of-the-art performance for the classification task, making them useful probes for the information content in the jets. \rev{However, due to the high-dimensionality of their inputs and the low-level nature of the data, it could be challenging to directly interpret low-level models in an experimental context. Under an experimental setting, having a smaller number of well-understood high-level inputs which capture the necessary information would be preferred. Thus, we aim to identify new high-level observables which can be added to the existing observables to close the performance gap.}

A strategy for identifying jet observables which are able to bridge the performance gap is described in Ref.~\cite{Faucett:2020vbu}, which searches among the pool of observables called energy flow polynomials (EFPs)~\cite{Komiske:2017aww} to identify those that yield similar classification decisions as DNNs trained on low-level detector data. This strategy, however, applies only to binary decision functions~\cite{Collado:2020fwm,Collado:2021opi}. We leave a generalization of that method to multi-class networks for future work, and instead apply a simpler but commonly-used technique for variable selection.  First, we employ a large set of EFPs in combination with the existing $\textrm{N}$-subjettiness variables to capture the information needed to match the performance of the Transformer and the PFN. Then, we systematically reduce the set of observables to find the minimum set that best approximates the performance of the constituent-based networks. 

\subsection{Adding Energy-Flow Polynomials}

As described in~\cite{Komiske:2017aww}, EFP observables are constructed as nested sums over jet constituents transverse energy, or equivalently $\pT$,  scaled by their pairwise angular separation $\theta_{ij}$. These parametric sums are described as the set of all isomorphic multigraphs, where for a jet with $M$ constituents:

\begin{align}
	\text{each node} &\Rightarrow \sum_{i = 1}^M z_i, \label{eq:EFP_node}  \\
	\text{each edge connecting nodes $k$ and $l$} &\Rightarrow \left(\theta_{i_k i_j}\right) \label{eq:EFP_edge} . 
\end{align}

Each graph can be further parameterized by the weighing factors $(\kappa, \beta)$, where

\begin{align}
	(z_i)^\kappa &= \left(\frac{p_{\textrm{T}i}}{\sum_j p_{\textrm{T}j}} \right)^\kappa, \label{eq:EFP_z} \\
	\theta^\beta_{ij} &= \left(\Delta \eta_{ij}^2 + \Delta \phi_{ij}^2 \right)^{\beta/2}. \label{eq:EFP_theta}
\end{align}

Here, $p_{\textrm{T}i}$ is the transverse momentum of constituent $i$, and $\Delta \eta_{ij}$ ($\Delta \phi_{ij}$) is the pseudorapidity (azimuthal) difference between constituents $i$ and $j$. Following the documentation of~\cite{Komiske:2017aww}, each EFP is accompanied by its unique identifier ($n$, $d$, $k$), which specifies the number of nodes $n$, edges $d$, and index $k$ of the corresponding graph. 

\begin{figure}[!h]
    \centering
    \includegraphics[width=\linewidth]{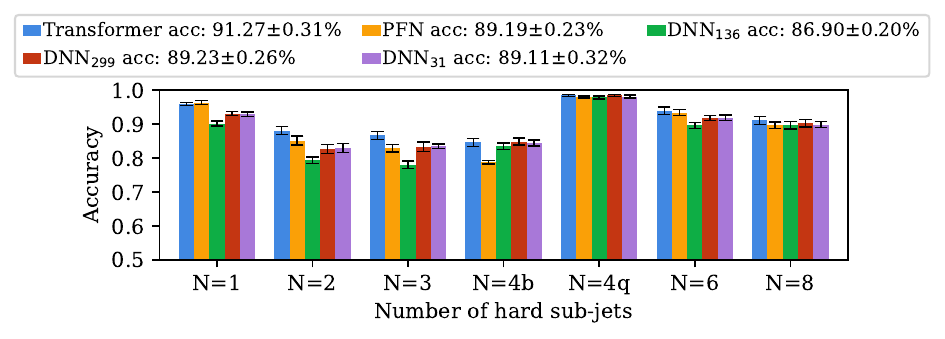}
    \caption{
        \rev{Mean 10-fold} accuracy and statistical uncertainty of the various networks predictions for jets in each class.
        }
    \label{fig:acc_prong_efps}

    \centering
    \includegraphics[width=\linewidth]{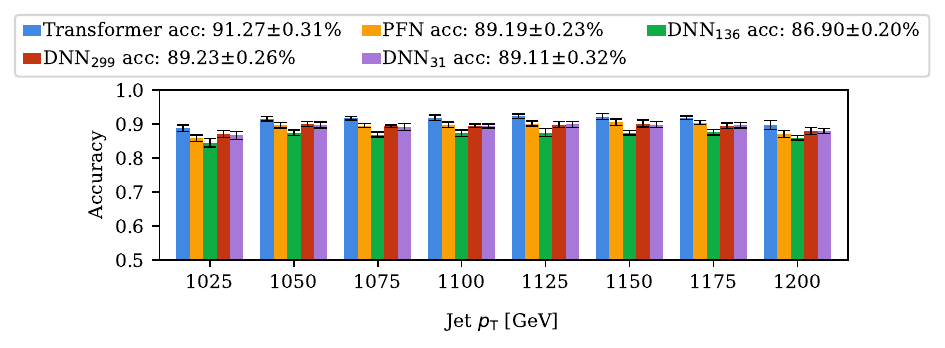}
    \caption{
     \rev{Mean 10-fold} accuracy and statistical uncertainty of the various networks predictions for jets within ranges of jet \pT.
     \rev{The jets are binned according to their \pT\ in intervals of 25 GeV. The x-axis labels correspond to the upper bound of the intervals.}
     }
    \label{fig:acc_prong_efps_pT}

    \centering
    \includegraphics[width=\linewidth]{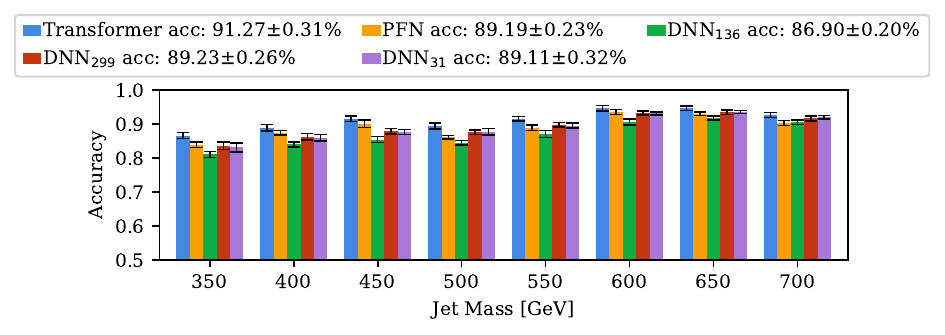}
    \caption{
     \rev{Mean 10-fold} accuracy and statistical uncertainty of the various networks predictions for jets within ranges of jet mass.
     \rev{The jets are binned according to their mass in intervals of 50 GeV. The x-axis labels correspond to the upper bound of the intervals.}
     }
    \label{fig:acc_prong_efps_mass}
\end{figure}

For our studies, we select all connected (prime) EFP graphs with five or fewer edges and weighing factors $\kappa = 1$ and $\beta \in \{ \tfrac{1}{2}, 1, 2 \}$, for a total of 162 observables, which are denoted by EFP$^{\beta}(n,d,k)$ in what follows. We also include the constituent multiplicity, which is described by the one-node EFP with $\kappa = 0$, and has been shown to be a useful observable for jet discrimination~\cite{gallicchio2013}. Combined with the $\textrm{N}$-subjettiness variables and the jet mass, the new augmented dataset has 299 observables.  We train a dense network on the 299 observables, labelled DNN$_{299}$, with the same hyperparameters as DNN$_{136}$. The resulting overall accuracy is $89.23\%$, as seen in Figure~\ref{fig:acc_prong_efps}. With the augmented set, we are thus able to match the overall performance of the PFN, and to close the performance gap with the Transformer to approximately 2$\%$.  

The accuracies of DNN$_{299}$ across the jet \pT~and mass spectra are shown in Figure~\ref{fig:acc_prong_efps_pT}~ and Figure \ref{fig:acc_prong_efps_mass}, respectively. In most ranges, DNN$_{299}$ matches or closely approximates the performance of the PFN. A small but persistent gap remains between the DNN$_{299}$ and Transformer models, indicating that the Transformer model is still utilising useful information not available in the augmented observable set. Additional EFP observables may be able to further narrow the gap, such as EFPs with energy and angular measures not considered in this paper, or with a larger number of edges (degree), which allow for more complex polynomial forms. The costly computation of these variables makes this infeasible at the present time and so is left for future work.

\subsection{Selecting Energy-Flow Polynomials}

The next step in the selection process is to identify the minimal set of high-level observables which manages to close the gap with the PFN and approximate the Transformer.  The core idea of the observable selection strategy is to apply $L_{1}$ regularization, also known as LASSO~\cite{lasso} regularization, to zero out the weights of the observables that are unimportant to the network during training.  In principle, the regularization must be applied to the input weights of the DNN$_{299}$ network, which are given by a matrix of rank [800, 299], namely $W=[W_{1}, W_{2}, ..., W_{299}]$, where $W_{i}$ is a column vector of length 800. However, since we are interested in finding the minimal set of observables which are the most relevant for the classification task, we encourage the network to shrink the entire weight column vector of the irrelevant observables down to zero by implementing a learnable gate parameter $g_i$, such that $W_{i}'= [g_{i}w_{i,1}, ..., g_{i}w_{i,800}]$.  This allows us to apply the regularization on the gate parameters $\bm{g}=[g_1, g_2, ..., g_{299}]$ during training.  It can then be seen that when the LASSO regularization shrinks $g_{i}$ down to zero, the entire vector of $W_{i}'$ also shrinks to zero, and we can confidently exclude the $i$th observable during training.  The overall loss function can be written as
\begin{equation}
    L = -\log f(Y, Y_{\text{pred}}) + \lambda \cdot \sum_{i=1}^{299}|g_{i}|,
\end{equation}
where the first term $-\log f(Y, Y_{\textrm{}{pred}})$ is the negative log likelihood of the predicted label $Y_{\textrm{}{pred}}$ and true label $Y$, and the second term is the LASSO regularization term on the gate parameter $g$ with regularization strength parameter $\lambda$. 

\begin{figure}[!h]
    \centering
 	\includegraphics[width=0.9\textwidth]{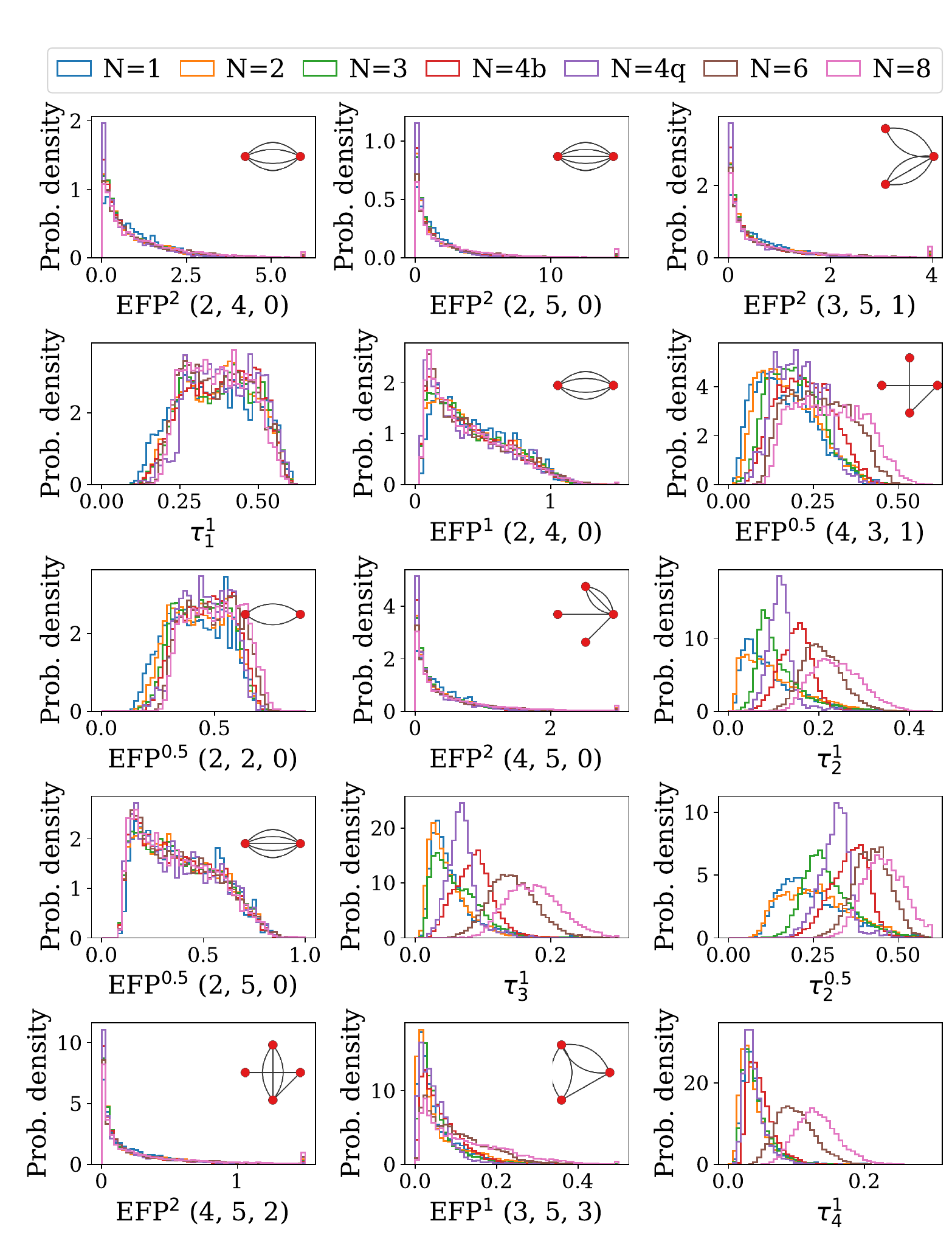}
    \caption{Distributions of the first half of the most important LASSO-selected observables.  The plots are ranked left-to-right top-to-bottom in order of importance according to the method described in Sec.~\ref{sec:ranking_obs}. For each EFP observable, the corresponding graph is also displayed for visualization purposes. The jet mass, which ranks between $\tau^1_3$ and $\tau_2^{0.5}$ as the 12th most important variable, is omitted for brevity as this distribution is approximately flat by design, and is already shown in Figure~\ref{fig:masspt}. The distributions of the rest of the LASSO-selected observables are shown in Figure~\ref{fig:selected_EFP_hist_part2}.
    }
    \label{fig:selected_EFP_hist_part1}
\end{figure}

\begin{figure}[!h]
    \centering
 	\includegraphics[width=0.9\textwidth]{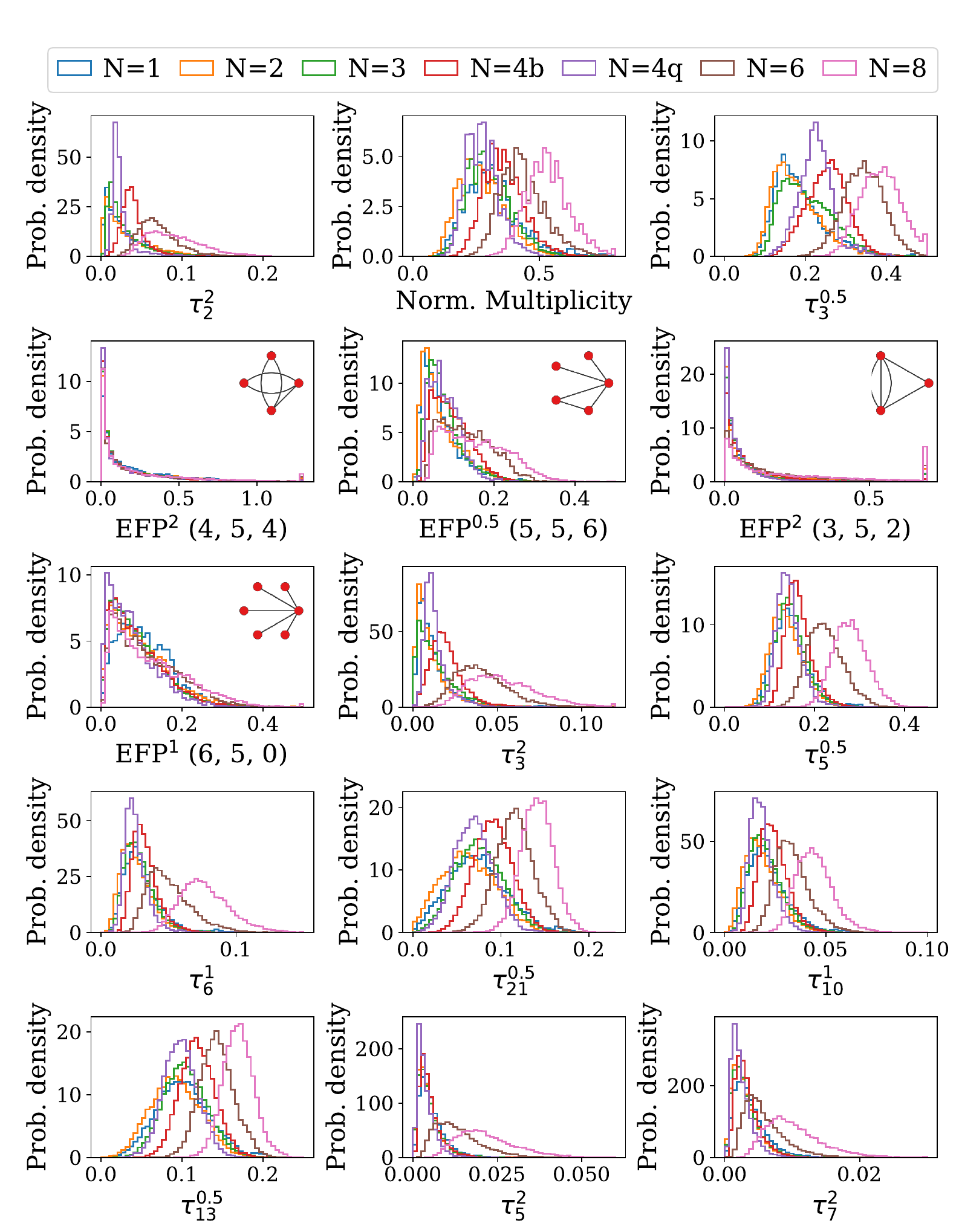}
    \caption{Distributions of the second half of the most important LASSO-selected observables. The plots are ranked left-to-right top-to-bottom in order of importance according to the method described in Sec.~\ref{sec:ranking_obs}. For each EFP observable, the corresponding graph is also displayed for visualization purposes.  The distributions of the rest of the LASSO-selected observables are shown in Figure~\ref{fig:selected_EFP_hist_part1}.
    }
    \vspace{0.5cm}
    \label{fig:selected_EFP_hist_part2}
\end{figure}

A grid search of the regularization strengths ranging from 1 to 10 indicates that the best regularization strength parameter value for the classification task is $\lambda = 5$. To further limit the minimum observable set, only observables with a gate parameter of $|g_{i}| > 0.01$ are kept.  With these settings, the selection strategy results in 31 LASSO-selected observables whose distributions are shown in Figures~\ref{fig:selected_EFP_hist_part1} and~\ref{fig:selected_EFP_hist_part2}.

We train a dense network operating on the 31 LASSO-selected observables, DNN$_{31}$. A hyperparameter search indicates that the best structure for the network has six hidden layers of size (800-800-800-800-800-32) and \textsc{ReLu}~\cite{inproceedings} activation function. Dropout with rate 0.3 and batch normalization are applied respectively after every hidden layer. Please refer to Table~\ref{tab:ModelSummaries} for the model summary of DNN$_{31}$.

The DNN$_{31}$ network achieves an overall accuracy of $89.11 \pm 0.32\%$, as shown in Figure~\ref{fig:acc_prong_efps}. Note that for all classes, the DNN$_{31}$ matches or closely approximates the performance of the DNN$_{299}$, indicating that the LASSO selection strategy has succeeded in identifying the observables which capture most of the information relevant to the classification task.\footnote{Although the LASSO-selected observables do a good job at matching the performance of the full 299 observable set, the overlapping nature of the EFP observables makes it likely that this is not a unique solution and that different settings in the LASSO selection process may yield a somewhat different subset of observables.}

Figure~\ref{fig:acc_prong_efps_pT} and Figure~\ref{fig:acc_prong_efps_mass} show the accuracies across the jet~$\pT$ and mass spectra, respectively. Both figures show similar trends for DNN$_{31}$ as for the other networks; uniform predictions across the jet~$\pT$ spectrum, and generally rising accuracy with jet mass.

\subsection{Importance of EFP observables}
\label{sec:ranking_obs}

After narrowing down the set of observables to 31 LASSO-selected observables, we attempt to interpret the strategies of the low-level networks in terms of the selected observables by ranking them in order of importance.

We measure how much the DNN$_{31}$ model relies on each of the observables to make a prediction by randomly shuffling them during test time.  That is, the $k$-th observable ($k \in \{1,\ldots,  31\}$) in the test set is shuffled by randomly replacing it with a corresponding observable value from the training set, on an event-by-event basis.  This maintains the marginal distribution for the $k$-th observable, but destroys any correlations with other observables.  The performance of the network is then evaluated on the set of events which includes the shuffled observable.  This method allows us to determine the importance of the $k$-th observable by measuring the drop in the performance of the network relative to the unshuffled set.  By this method, an observable is considered important if shuffling its values decreases the accuracy of the model. Likewise, an observable is considered unimportant if shuffling its values leaves the accuracy unchanged, as it indicates that the model does not heavily rely on that observable when making predictions.

Table~\ref{tab:observable_ranking} shows the ranking of the 31 LASSO-selected observables. The observables are ranked in order of importance, with the most important corresponding to the largest drop in accuracy of the model when shuffled.  These observables mainly consist of EFPs with four or less nodes, and $\textrm{N}$-subjettiness variables with $\textrm{N} < 8$.  The latter is not surprising as it is expected that $\textrm{N}$-subjettiness variables with $\textrm{N}$ significantly larger than the number of sub-jets are less important for the classification problem. 

\begin{table}[!h]
    \centering
    \makebox[0pt][c]{\parbox{1.2\textwidth}{
        \begin{minipage}[t]{0.45\hsize}
            \centering 
            \begin{tabular}[t]{ c | l | c }
                \hline\hline
                Rank & Observable & Accuracy drop \\
                \hline
                1 & EFP$^{2}$ (2, 4, 0)    & $50.60 \pm 1.95 \%$ \\
                2 & EFP$^{2}$ (2, 5, 0)    & $44.81 \pm 2.28 \%$ \\
                3 & EFP$^{2}$ (3, 5, 1)    & $41.68 \pm 2.35 \%$ \\
                4 & $\tau_1^1$             & $41.49 \pm 1.06 \%$ \\
                5 & EFP$^{1}$ (2, 4, 0)    & $38.81 \pm 1.41 \%$ \\
                6 & EFP$^{0.5}$ (4, 3, 1)  & $37.99 \pm 1.02 \%$ \\
                7 & EFP$^{0.5}$ (2, 2, 0)  & $37.26 \pm 1.38 \%$ \\
                8 & EFP$^{2}$ (4, 5, 0)    & $35.37 \pm 0.70 \%$ \\
                9 & $\tau_2^1$             & $34.97 \pm 0.63 \%$ \\
                10 & EFP$^{0.5}$ (2, 5, 0) & $33.66 \pm 2.02 \%$ \\
                11 & $\tau_3^1$            & $30.26 \pm 1.08 \%$ \\
                12 & Norm. Jet Mass        & $29.44 \pm 0.92 \%$ \\
                13 & $\tau_2^{0.5}$.       & $29.38 \pm 0.90 \%$ \\
                14 & EFP$^{2}$ (4, 5, 2)   & $27.66 \pm 1.66 \%$ \\
                15 & EFP$^{1}$ (3, 5, 3)   & $27.58 \pm 0.88 \%$ \\
                16 & $\tau_4^{1}$          & $26.74 \pm 0.96 \%$ \\
                \hline\hline
            \end{tabular}
        \end{minipage}
        \begin{minipage}[t]{0.45\hsize}
            \centering
            \begin{tabular}[t]{ c | l | c }
                \hline\hline
                Rank & Observable & Accuracy drop \\
                \hline
                17 & $\tau_2^{2}$           & $26.17 \pm 0.92 \%$ \\
                18 & Norm. Multiplicity     & $24.87 \pm 0.77 \%$ \\
                19 & $\tau_3^{0.5}$         & $24.53 \pm 0.68 \%$ \\
                20 & EFP$^{2}$ (4, 5, 4)    & $23.00 \pm 1.10 \%$ \\
                21 & EFP$^{0.5}$ (5, 5, 6)  & $22.59 \pm 1.07 \%$ \\
                22 & EFP$^{2}$ (3, 5, 2)    & $21.39 \pm 0.80 \%$ \\
                23 & EFP$^{1}$ (6, 5, 0)    & $19.82 \pm 1.09 \%$ \\
                24 & $\tau_3^{2}$           & $16.35 \pm 0.77 \%$ \\
                25 & $\tau_5^{0.5}$         & $15.00 \pm 0.85 \%$ \\
                26 & $\tau_6^{1}$.          & $13.14 \pm 0.69 \%$ \\
                27 & $\tau_{21}^{0.5}$      & $5.08 \pm 0.40 \%$ \\
                28 & $\tau_{10}^{1}$        & $4.94 \pm 0.37 \%$ \\
                29 & $\tau_{13}^{0.5} $     & $4.15 \pm 0.44 \%$ \\
                30 & $\tau_5^{2}$           & $2.75 \pm 0.34 \%$ \\
                31 & $\tau_7^{2}$           & $0.63 \pm 0.25 \%$ \\
                \hline\hline
            \end{tabular}
        \end{minipage}
    }}
    \caption{Ranking of the 31 LASSO-selected observables in order of importance, with the most important corresponding to the largest drop in accuracy of the model. The corresponding distributions and EFP graphs are displayed in Figures~\ref{fig:selected_EFP_hist_part1} and~\ref{fig:selected_EFP_hist_part2}.
    }
    \label{tab:observable_ranking}
\end{table}

\begin{figure}[!h]
    \centering
    \includegraphics[width=0.9\linewidth]{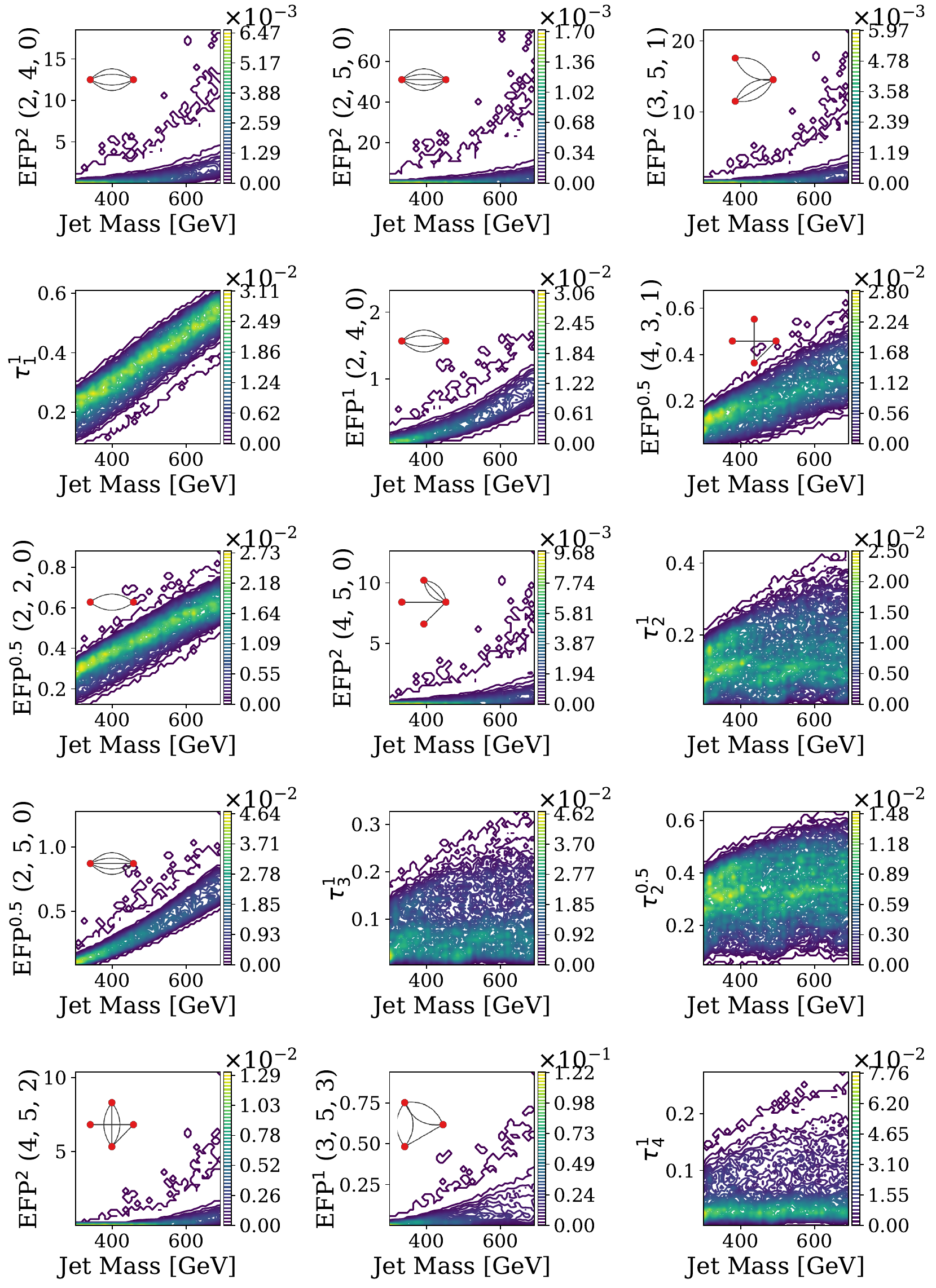}
    \caption{
     Contour plots showing the probability density of the first half of the most important LASSO-selected observables versus the jet mass. The plots are ranked in order of importance, excluding the normalized jet mass, which ranks at number 12 and is omitted for brevity. The contour plots of the rest of the LASSO-selected observables are shown in Figure~\ref{fig:ranked_obs_vs_mass_part2}.
     }
    \label{fig:ranked_obs_vs_mass_part1}
\end{figure}

\begin{figure}[!h]
    \centering
    \includegraphics[width=0.9\linewidth]{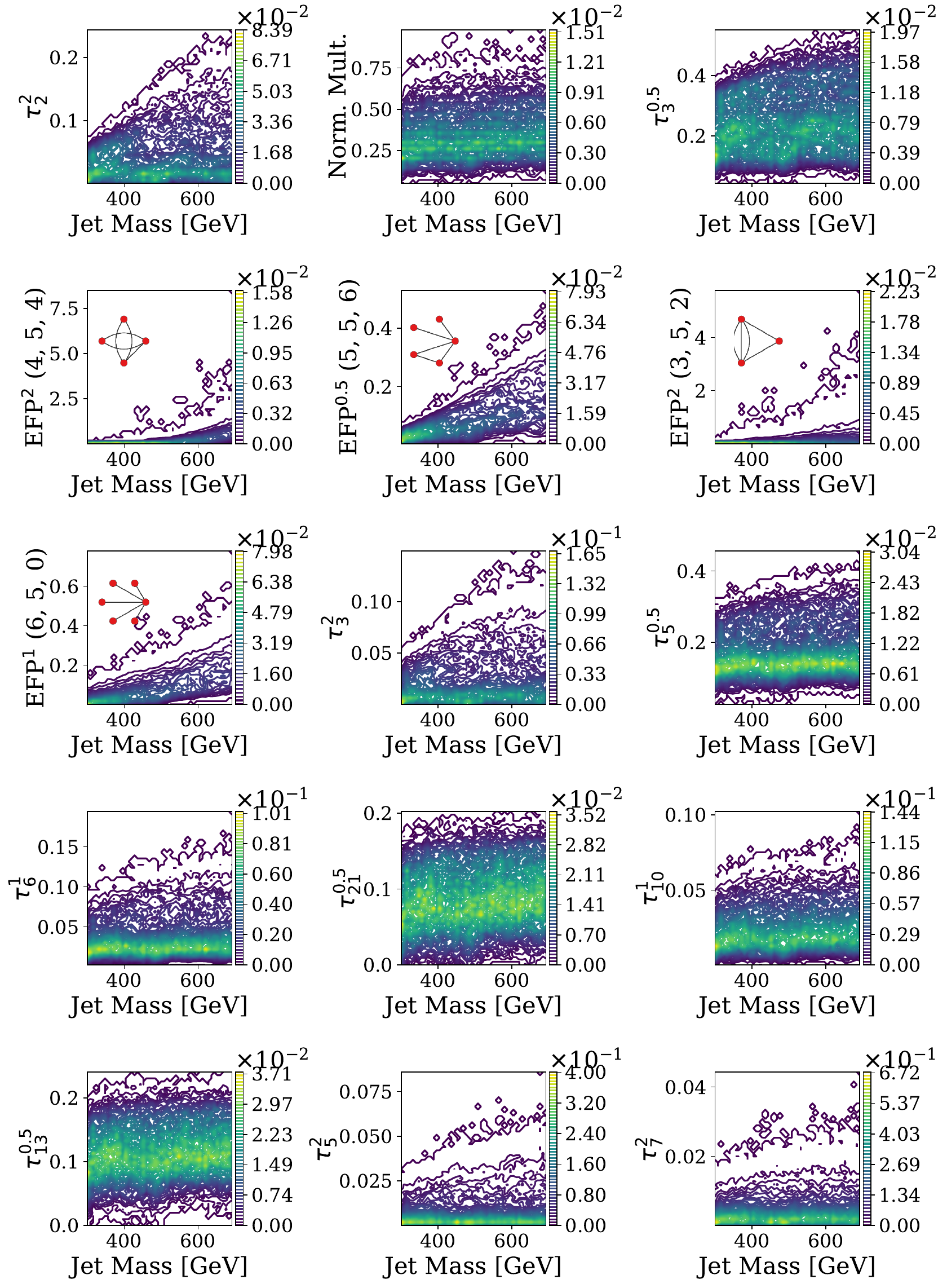}
    \caption{
     Contour plots showing the probability density of the second half of the most important LASSO-selected observables versus the jet mass. The contour plots of the rest of the LASSO-selected observables are shown in Figure~\ref{fig:ranked_obs_vs_mass_part1}.
     }
     \vspace{1cm}
    \label{fig:ranked_obs_vs_mass_part2}
\end{figure}

Figures~\ref{fig:ranked_obs_vs_mass_part1} and~\ref{fig:ranked_obs_vs_mass_part2} show contour plots of the LASSO-selected observables versus the jet mass.  The contour plots indicate that there is a correlation between the top ranked observables and the jet mass.  This correlation is particularly striking for the $\tau_{1}^1$ variable, which is the fourth\footnote{An earlier version of this paper had identified $\tau_{1}^1$ as the third most important observable.} most important observable and can be interpreted as a measure of how collimated the jet constituents are along the one $\textrm{N}$-sub-jet axis, with lower $\tau_{1}^1$ values indicating that the constituents are more collimated. By inspection of this observable, we can explain the lower classification power of the networks at lower jet mass values from Figure~\ref{fig:acc_prong_efps_mass}, as jets with lower jet masses may be more collimated and thus harder to classify.

\begin{figure}[!h]
    \centering
    \includegraphics[width=0.87\linewidth]{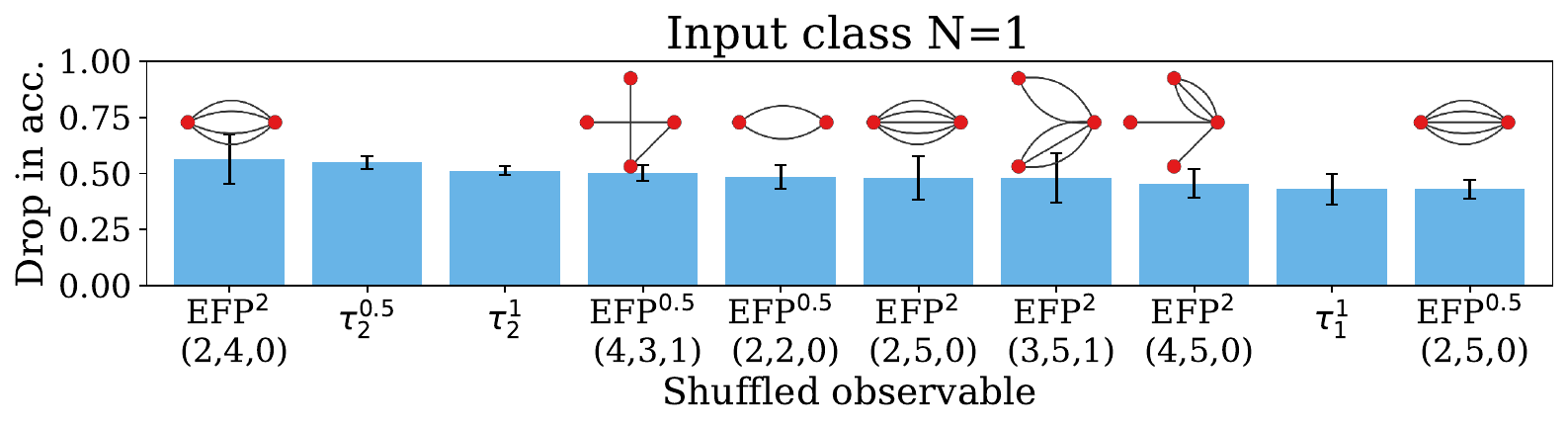}
    
    
    \includegraphics[width=0.87\linewidth]{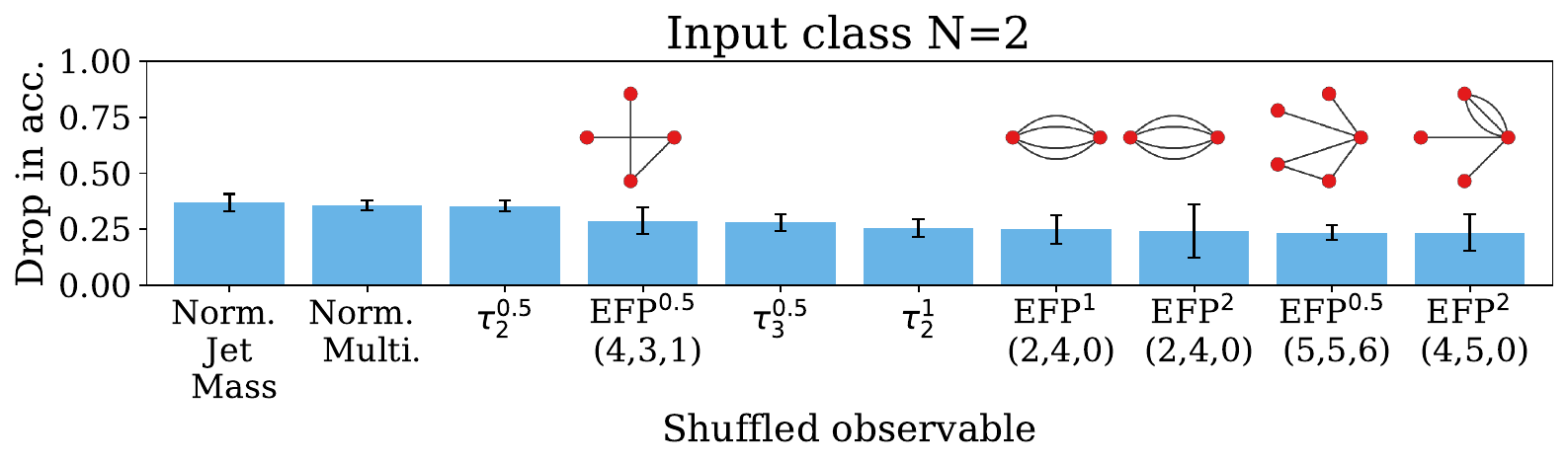}
    
    
    \includegraphics[width=0.87\linewidth]{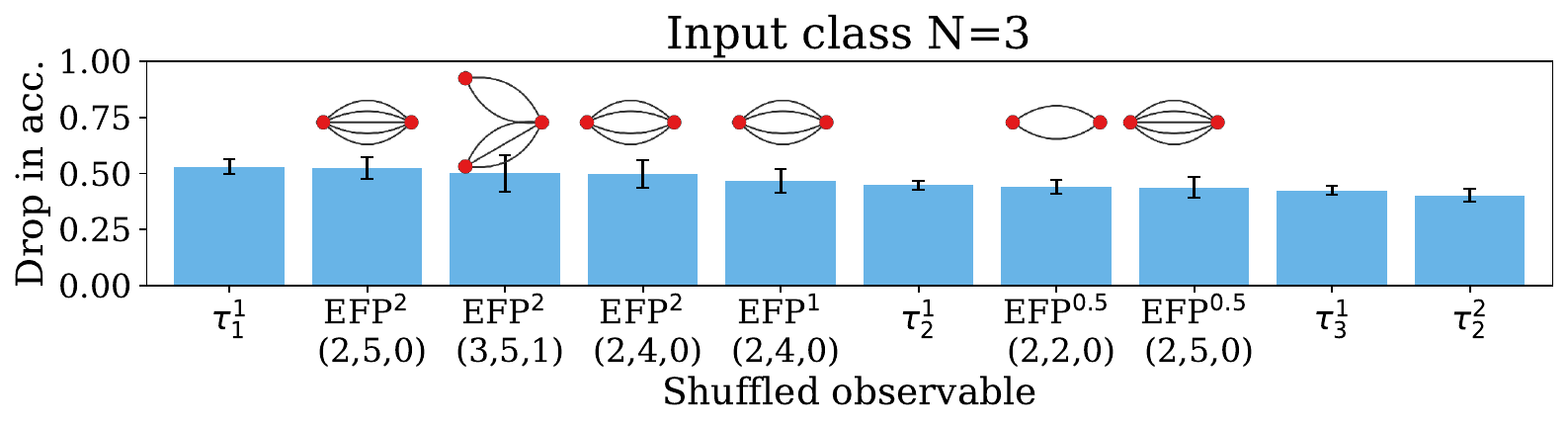}
    
    
    \includegraphics[width=0.87\linewidth]{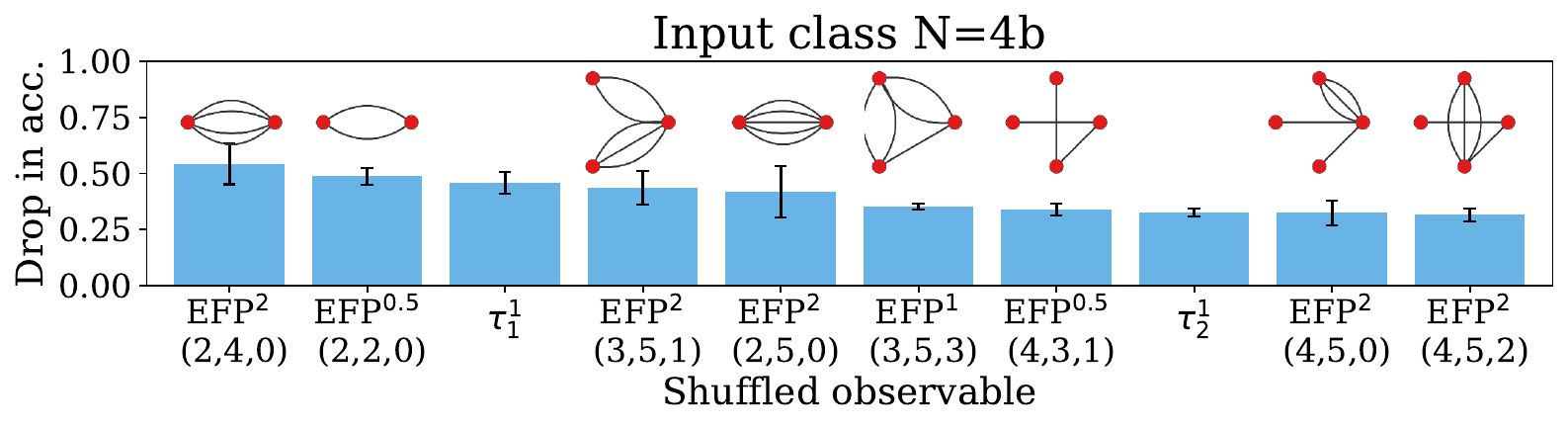}
    
    
    \includegraphics[width=0.87\linewidth]{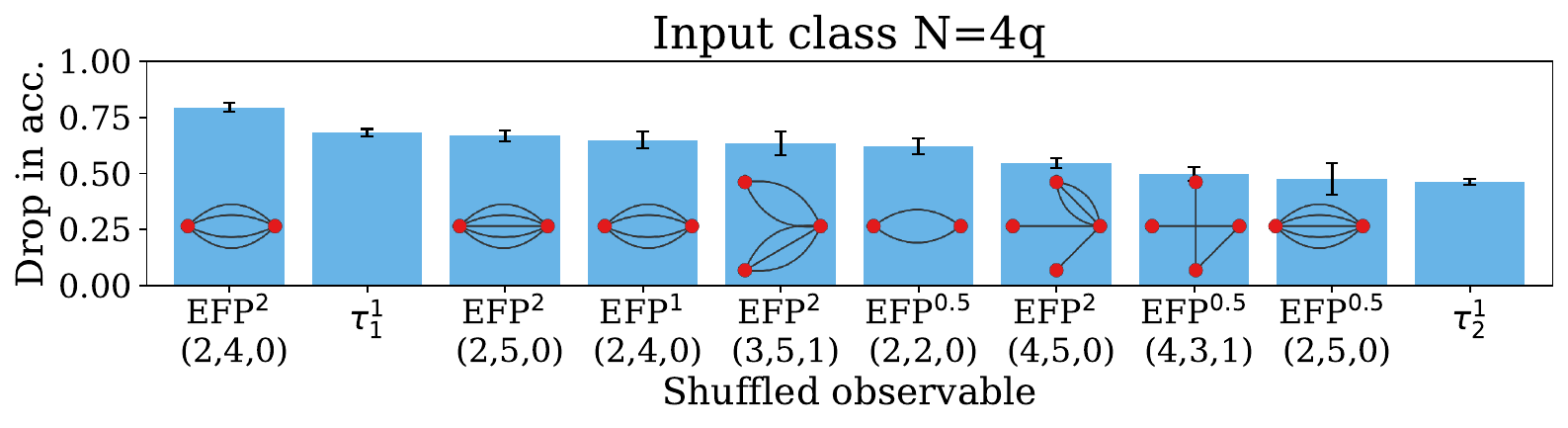}

    \caption{ 
    Accuracy drop and statistical uncertainty of the top 10 observables per input class $N=1, 2, 3, 4b, 4q$. The observables are ranked in order of importance, where the most important are those resulting in the largest drops in accuracy of the model when randomly replaced with observable values from the training set.  The EFP graphs are included for visualization purposes.}
    \label{fig:efp_analysis_n1_to_3}
\end{figure}

\begin{figure}[htb]
    \centering

    \includegraphics[width=0.87\linewidth]{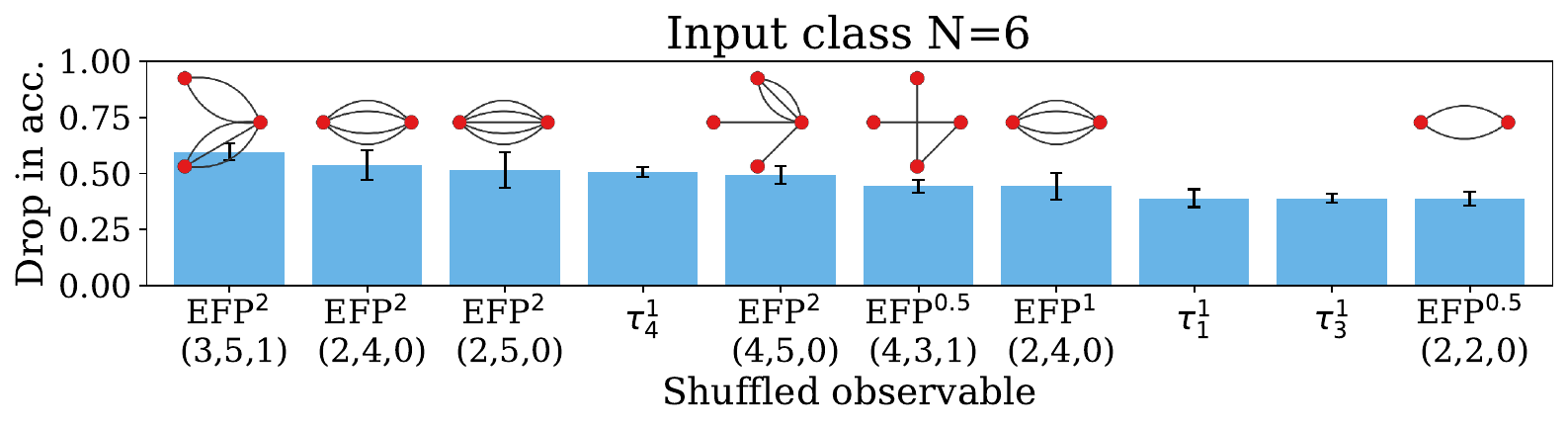}
    
    
    \includegraphics[width=0.87\linewidth]{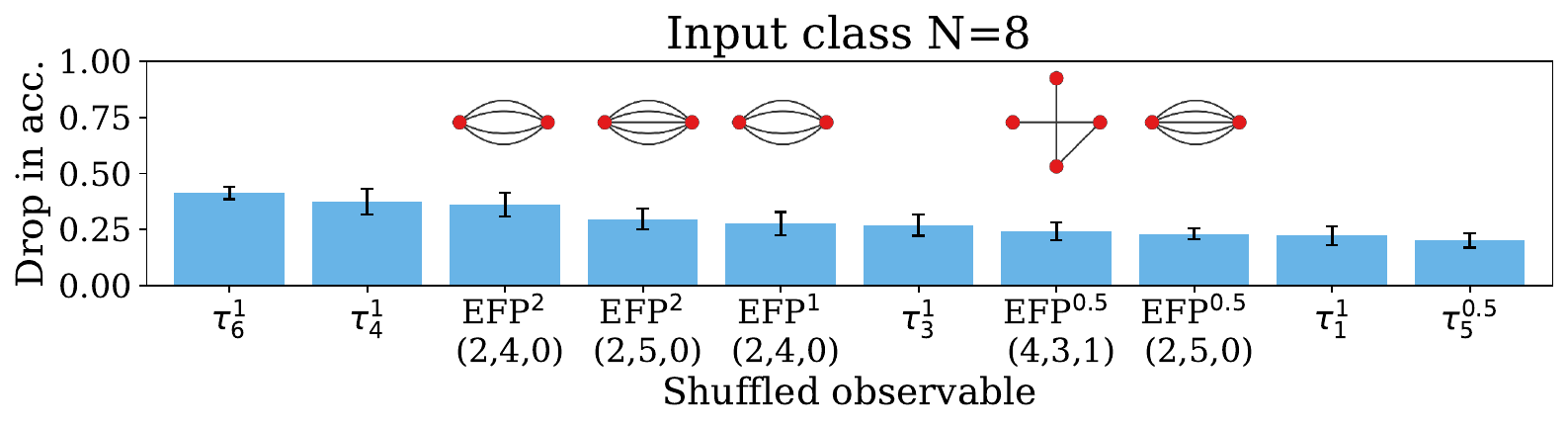}
    
    \caption{Accuracy drop and statistical uncertainty of the top 10 observables per input class $N=6, 8$. The observables are ranked in order of importance, where the most important are those resulting in the largest drops in accuracy of the model when randomly replaced with observable values from the training set.  The EFP graphs are included for visualization purposes.}
    \label{fig:efp_analysis_n4_to_8}
\end{figure}

As the importance of an observable may vary across the several classes, we measure how much each of the classes relies on each of the 31 LASSO-selected observables by measuring the drop in class accuracy as the observables are shuffled.  Figures~\ref{fig:efp_analysis_n1_to_3} and \ref{fig:efp_analysis_n4_to_8} show the drop in class accuracy of the top 10 observables per class. For the $N = 1, 2, 3, 4b, 4q$ classes, the top observables mainly consist of EFPs with four or less nodes, and $\textrm{N}$-subjettiness variables with three or less sub-jet axes. It is interesting to note that the constituent multiplicity and jet mass are highly important for the $N = 2$ class. It is also interesting how although the classes $N = 4b, 4q$ share some of the top observables, the relative importance of these observables is different for both classes. 
For the $N = 6, 8$ classes, we find that while EFPs with with four or less nodes are still some of the most important observables, the $\textrm{N}$-subjettiness variables with $\textrm{N}$ larger than 3 become more relevant. Particularly, for the $N = 8$ case, $\tau_6^1$ and $\tau_4^1$ are the two most important observables, indicating that, at large scales, these jets may look like jets with six or four sub-jets. 

For most cases, the most relevant observables are two-node EFPs, or $\textrm{N}$-subjettiness variables with relatively small $\textrm{N}$ ($\textrm{N} = $1, 2, 3, or 6). Subsequent observables generally consist of EFPs with more complex shapes or $\textrm{N}$-subjettiness variables with larger $\textrm{N}$. This suggests that the network may be focusing on distinguishing between the different classes of jets by first utilizing observables that broadly capture the number of sub-jets, and later utilizing more complex observables that specialize on capturing specific traits for each class, such as the presence of collimated jets or subtle differences in the topologies of the jets. The latter is further explored in Sec.~\ref{sec:4b4q}.

\subsection{Topology Dependence}
\label{sec:4b4q}

The networks have learned to distinguish jets with various numbers of hard sub-jets. We do not, however, claim that they have learned to identify \emph{any} jet with this number of hard sub-jets. On the contrary, it is likely that the patterns of energy depositions depend on the details of the topology, such as the invariant mass of intermediate resonances and the jet flavors. The networks' ability to distinguish jets from $G\rightarrow HH\rightarrow 4b$ and $G\rightarrow WW\rightarrow 4q$ is an example. In this section, we explore the dependence of the networks' classification strategies on these details of the sub-jet topology.

We generate two additional samples of jets with $N=4$ hard sub-jets. In the first, labeled $4bM_W$, jets are produced with the $G\rightarrow HH\rightarrow 4b$ process, but the mass of the Higgs boson has been set to the mass of the $W$ boson to more closely align with the $4q$ sample.  In the second, labeled $4qM_H$, jets are produced with the $G\rightarrow WW\rightarrow 4q$ process, but the mass of the $W$ boson mass has been set to the mass of the Higgs boson to more closely align with the $4b$ sample.  As with other samples, the $4bM_W$ and $4qM_H$ samples are selected such that they have uniform distributions in jet mass and \pT.

\begin{figure}[!h]
    \centering
    \includegraphics[width=\linewidth]{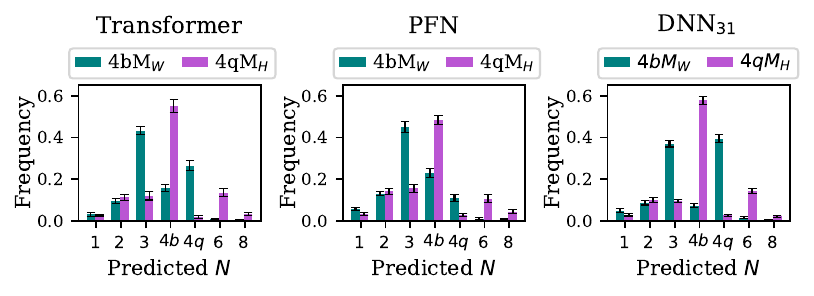}
    \caption{
    Frequency of class identification on samples with modified topology ($4bM_W$ and $4qM_H$) for networks trained with the standard topologies ($N = 1, 2, 3, 4b, 4q, 6, 8$).}
    \label{fig:4prong_perf}
\end{figure}

We evaluate the networks' predictions on the $4bM_W$ and $4qM_H$ samples. The frequency of classification outputs of these two samples are shown in Figure~\ref{fig:4prong_perf}. In the $4qM_H$ case, all three networks mainly classify the samples as $N=4b$ jets, and rarely classify them as $N=4q$ jets. This suggests a strong correlation between the intermediate masses and the final state kinematics which the networks are learning. Conversely, in the $4bM_W$ case, the high- and low-level networks result in different classification predictions. The low-level networks mainly classify the $4bM_W$ samples as $N=3$ jets, while the DNN$_{31}$ mainly classifies these samples as $N=4q$ or $N=3$. The more frequent prediction of $N=3$ by the low-level networks hints at the possibility that these networks are identifying complex details of the jet topology, which goes beyond simply identifying hard sub-jets, as the $N=3$ sample includes an intermediate $W$ boson as well as a $b$-jet.

The different predictions of the networks come to no surprise as, although they reach similar accuracies, they each have different learning strategies. The DNN$_{31}$ learns functions of the 31 LASSO-selected observables, while the PFN learns an event-level latent representation of the jets by summing over the constituents, and the Transformer uses self-attention mechanisms to determine which parts of the constituent sequence it should focus on when making predictions.\footnote{We analyzed the redundancy of the networks' strategies by computing the boost in performance obtained by concatenating their outputs. The results suggested that the networks are using unique information, as concatenating any two networks resulted in a mild boost in performance. Further investigation of the nature of this unique information was reserved for future work.} It is interesting to note that despite their different learning strategies, all taggers seem to be dependent on the detailed topologies, which is a benefit rather than a flaw in studies aiming to classify specific sub-jet topologies.

\section{Conclusions}
\label{sec:conc}

We have studied the task of classifying jets with a large number of hard sub-jets (up to $N=8$), comparing the performance of networks trained on theoretically-motivated high-level observables to networks operating on low-level jet constituents.  We find that the networks trained on the jet mass and high-level $\textrm{N}$-subjettiness variables are able to effectively discriminate between all classes, particularly for jets with many hard sub-jets or large jet mass. \rev{These findings prove useful for current studies searching to classify jets with multiple hard sub-jets. In addition, our results bode well for future jet studies in high energy collider settings, where jets with additional hard sub-jets will become more important as the high-luminosity LHC collects large datasets in which high-$p_\textrm{T}$ objects appear in greater numbers~\cite{CMS:2021beq,Aad:2020ddw}.}

However, the networks trained on the $\textrm{N}$-subjettiness observables fall somewhat short in classification accuracy compared to networks utilising low-level constituent information, particularly for jets with fewer sub-jets or smaller jet mass.  We thus supplement these observables with a large set of EFPs, for a total of 299 high-level observables.  The network trained on the augmented high-level observable set is then able to match the performance of the well studied PFN networks, and to approximate the performance of the Transformer model.  The performance gap between the Transformer model and the high-level observables is small but significant, suggesting that there remains useful discriminating information in the low-level constituents that is not captured even by the full set of 299 input high-level observables. 

Utilising LASSO regularization for feature selection, we find the most important observables which contain most of the information relevant for the classification task. We identify 31 high-level observables which largely bridge the gap to the PFN, and rank them in order of importance to gain some insights into the nature of the learning strategies of the networks. We find that the most important observables are mainly EFPs with four or less nodes, and $\textrm{N}$-subjettiness variables with $\textrm{N} < 8$.  We also identify the constituent multiplicity as one of the top observables, and find it to be particularly important for classifying $N = 2$ jets.

By analyzing the most important observables for each input class $N$, we find that the strategy of the networks may rely on first utilizing simple observables that broadly capture the number of sub-jets, such as EFP polynomials with only a few nodes or $\textrm{N}$-subjettiness variables with small $\textrm{N}$ values. Subsequently, the networks may utilize more complex observables to capture subtle traits of the topology of the jets. This is further confirmed in our results, which reveal the classifiers to have a strong topology dependence, as they appear to be sensitive to the sub-jet resonance masses and flavor rather than simply being sensitive to the sub-jet multiplicity. Future work may involve disentangling the nature of this information to probe more deeply the classification strategy for high-multiplicity sub-jets.

\acknowledgments

The authors are grateful to Jesse Thaler, Andrew Larkoski, Tilman Plehn, Gregor Kasieczka, Ben Nachman, and Joakim Olsson for useful conversations and commentary on an earlier version of this paper. MF and DW are supported by the U.S.\ Department of Energy (DOE), Office of Science under Grant No. DE-SC0009920. AR is supported by the National Science Foundation under Grant No. 1633631.

\appendix

\section{Binary Classification}

A 7-class output network like the ones used in this paper can efficiently distinguish among the different types of jets. A more realistic case however is the separation of jets with many hard sub-jets from the overwhelming background of $N=1$ jets from QCD production. Here, we reintepret the 7-dimensional network output vector $\bm{\alpha}$ to perform six binary classification tasks.

\begin{figure}[!h]
    \centering
    \includegraphics[width=0.9\linewidth]{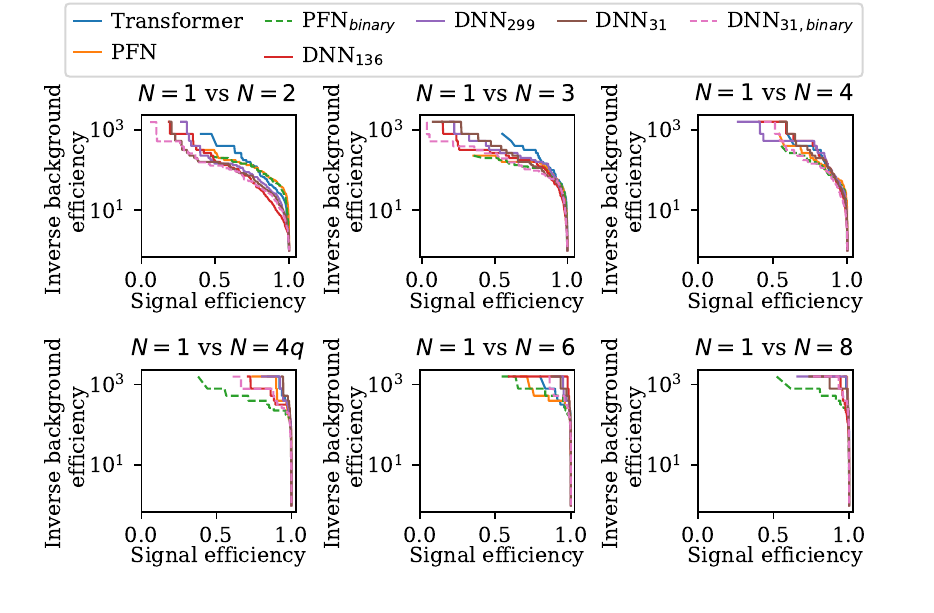}
    \caption{\rev{ROC curves of the various network studies classifying jets with $N=k$ versus $N=1$ hard sub-jets. See Table~\ref{tab:binary} for the ROC-AUC percentage scores.}}
    \label{fig:binary_ROCs}
\end{figure}

For discrimination between $N=1$ and $N=k$ sub-jets classes, we define the \textit{decision contrast score}:

\begin{align}
    C_{k} = \frac{\bm{\alpha}_{k}}{\bm{\alpha}_{k} + \bm{\alpha}_{1}}
\end{align}

\noindent where $C_{k}$ tends to 1 if the jet resembles the $N=k$ sub-jet class and tends to zero if it resembles the $N=1$ sub-jet class.  The binary classification performance can then be measured using the area under the receiver operating curve (ROC-AUC); see Table~\ref{tab:binary} for the ROC-AUC scores \rev{and Figure~\ref{fig:binary_ROCs} for the ROC curves.}

\begin{table}[ht]
    \centering
    \begin{tabular}{lcccc}
        \hline \hline 
            Model & $N=2$ & $N=3$ & $N=4b$ \\
        \hline
    Transformer & 
    99.06 $\pm$ 0.26& 
    99.58 $\pm$ 0.11& 
    99.66 $\pm$ 0.11& \\
    PFN &
    99.19 $\pm$ 0.13& 
    99.31 $\pm$ 0.14& 
    99.57 $\pm$ 0.12& \\
    \textbf{PFN$_{binary}$} &
    \textbf{98.84 $\pm$ 0.25}& 
    \textbf{99.01 $\pm$ 0.20}& 
    \textbf{99.29 $\pm$ 0.17}& \\
    DNN$_{136}$ & 
    97.04 $\pm$ 0.28& 
    98.97 $\pm$ 0.16& 
    99.39 $\pm$ 0.076& \\
    DNN$_{299}$ & 
    98.07 $\pm$ 0.23& 
    99.36 $\pm$ 0.17& 
    96.61 $\pm$ 0.063& \\
    DNN$_{31}$  & 
    98.02 $\pm$ 0.22& 
    99.31 $\pm$ 0.16& 
    99.57 $\pm$ 0.068& \\
    \textbf{DNN$_{31, binary}$}  & 
    \textbf{97.74 $\pm$ 0.28}& 
    \textbf{99.03 $\pm$ 0.13}& 
    \textbf{99.33 $\pm$ 0.11}& \\
            \hline \hline 
    \end{tabular}
    \vspace{1cm}

    \begin{tabular}{lccc}
        \hline \hline 
                Model & $N=4q$ & $N=6$ & $N=8$ \\
        \hline
    Transformer & 
    99.98 $\pm$ 0.021& 
    99.95 $\pm$ 0.033& 
    99.95 $\pm$ 0.047\\
    PFN &
    99.93 $\pm$ 0.039& 
    99.91 $\pm$ 0.048& 
    99.93 $\pm$ 0.034\\
    \textbf{PFN$_{binary}$}&
    \textbf{99.79 $\pm$ 0.079}& 
    \textbf{99.87 $\pm$ 0.052}& 
    \textbf{99.84 $\pm$ 0.12}\\
    DNN$_{136}$ & 
    99.96 $\pm$ 0.016& 
    99.92 $\pm$ 0.038& 
    99.96 $\pm$ 0.021\\
    DNN$_{299}$ & 
    99.98 $\pm$ 0.014& 
    99.96 $\pm$ 0.026& 
    99.98 $\pm$ 0.014\\
    DNN$_{31}$  & 
    99.97 $\pm$ 0.018& 
    99.95 $\pm$ 0.036& 
    99.97 $\pm$ 0.021\\
    \textbf{DNN$_{31, binary}$}  & 
    \textbf{99.91 $\pm$ 0.044}& 
    \textbf{99.95 $\pm$ 0.034}& 
    \textbf{99.92 $\pm$ 0.034}\\
            \hline \hline 
    \end{tabular}
    \caption{ROC-AUC percentage score for classifying jets with $N=k$ versus $N=1$ hard sub-jets, averaged over results from 10-fold cross validation. 
    }
    \label{tab:binary}
\end{table}

The binary classification results agree with the confusion matrices in Figure~\ref{fig:cf_matrix}, which show a larger degree of confusion between the $N=1$ and the $N=2, 3, 4b$ classes, and almost perfect classification between the $N=1$ and the $N=4q, 6, 8$ classes. 

As a cross check, the PFN and DNN$_{31}$ are retrained to perform the binary classification tasks. The results are included in Table~\ref{tab:binary} as PFN$_{binary}$ and DNN$_{31, binary}$. The multi-class taggers slightly outperform the binary taggers, particularly for lower $N$. This suggests that training on the different $N$-subjet classes may have resulted in a more robust decision boundary for the $N=1$ class. This hypothesis aligns with the results in~\cite{Hoffmann,Luo}, where training on sub-classes helped improve multi-class classification performance.

\section{Additional Technical Details}
\label{sec:model_details}

In this appendix, we describe the details of the machine learning models and network architectures.  The Transformer and PFN models are trained on the three-momenta of the simulated jet constituents, which are preprocessed normalizing the jet \pT{} to unity and subtracting the \pT-weighted angular means, as described in Sec.~\ref{sec:data}.  The dense networks are trained on high-level observables, which are strict functions of the preprocessed constituent three-momenta. This ensures that the dense networks have access to only a subset of the information available in the low-level jet constituents. 

The 10-fold average accuracy of the models and the number of parameters are summarized in Table~\ref{tab:ModelSummaries}.  Common properties across all networks include \textsc{ReLu}~\cite{inproceedings} activation functions for all hidden layers, and 7-dimensional softmax output functions to classify between all seven sub-jet classes.  All networks are trained using the Adam~\cite{Adam2014} optimizer for up to 1000 epochs, and with batch size of 256.

All networks were optimized by a hyperparameter search using the Sherpa~\cite{hertel2020sherpa} hyperparameter optimization library, while ensuring that the range of trainable parameters in the optimization is roughly the same for all networks.\footnote{PFN, Transformers, and DNN$_{136}$ models with a larger number of parameters were also considered resulting in slightly better accuracy values, but these networks were excluded from the study as their increased performances were only marginal when compared to the large computational cost of training the larger models.} The hyperparameters of the dense networks were optimized in the ranges: intermediate dimension $[600, 800]$, dropout ratio $[0.3, 0.4]$, and learning rate $[10^{-3}, 10^{-4}]$. The hyperparameters of the PFN were optimized in the ranges: $\phi$-module dimension $[128, 1024]$, $F$-module dimension $[128, 2014]$, $F$-module drop out rate $[0.1, 0.2]$, and learning rate $[10^{-3}, 10^{-4]}$. The hyperparameters of the Transformer were optimized in the ranges: hidden layer size $[256, 512]$, number of layers $[4, 8]$, and learning rate $[10^{-3}, 10^{-4}]$.

{\renewcommand{\arraystretch}{1.2}
\begin{table*}
\caption{\label{tab:ModelSummaries} 
Summary of the machine learning models used in the classification task. The table shows a brief description of each of the models, as well as the number of trainable parameters and the accuracy measured using 10-fold cross validation.}
\begin{tabular}{llcc}
\hline
\hline
Model & Description & No. of Params. & Accuracy\\ 
\hline
\hline
Transformer &
\makecell[l]{Transformer Network trained on\\ 
the jet constituents.} &
1,388,807 & 
91.27 $\pm$ 0.31 \%
\\
\hline
PFN &
\makecell[l]{Particle-Flow Network trained on\\ 
the jet constituents.} &
1,205,895 & 
89.19 $\pm$ 0.23 \%
\\
\hline
DNN$_{136}$ &
\makecell[l]{Fully-connected neural network\\ 
trained on the 135 N-subjettiness\\ 
observables and the norm. jet mass.} & 
2,732,519 &
86.90 $\pm$ 0.20 \%
\\
\hline
DNN$_{299}$ &
\makecell[l]{Fully-connected neural network \\
trained on the 135 N-subjettiness,\\
observables the normalized jet mass,\\
and the full set of EFP observables.} & 
2,862,919 &
89.23 $\pm$ 0.26 \%
\\
\hline
DNN$_{31}$ &
\makecell[l]{Fully-connected neural network \\
trained on the 31 LASSO-selected
\\observables.} & 
2,622,663 &
89.11 $\pm$ 0.32 \%
\\
\hline \hline
\end{tabular}
\end{table*}
}

Training of the Transformer and all dense networks was implemented in PyTorch~\cite{NEURIPS2019_9015} using NVIDIA V100 GPUs. Training of the PFN was implemented in Keras~\cite{chollet2015keras} with Tensorflow~\cite{tensorflow2015} backend using an NVIDIA RTX 2080 Ti GPU. The accuracy of the networks was checked to be insensitive to the deep learning library and GPU. Using the NVIDIA RTX 2080 Ti GPU as a benchmark, the average inference time for 256 samples was calculated to be 2.46$\pm$0.12 ms for the Transformer; 2.90$\pm$0.36 ms for the PFN; 0.58$\pm$0.01 ms for the DNN$_{136}$; 0.59$\pm$0.02 ms for the DNN$_{299}$; and 0.60$\pm$0.03 ms for the DNN$_{31}$. The inference times are provided to give a sense of scale of the latency of each algorithm, as we did not perform any inference optimization. 

\bibliographystyle{JHEP}
\bibliography{ref} 
\end{document}